\documentclass[12pt]{article}

\usepackage[a4paper,margin=2cm]{geometry}
\usepackage[utf8x]{inputenc}
\usepackage{amsmath,amsthm,amsfonts,amssymb,physics,color,mathtools,hyperref,url,
  qcircuit,pgfplots,tikz,stmaryrd,qcircuit}
\usetikzlibrary{quotes,shapes.geometric,arrows,fit,positioning,arrows.meta}
\hypersetup{colorlinks=true,citecolor=blue,urlcolor=cyan}
\usepackage[initials]{amsrefs} % abbreviates the first names of the authors all consistently
\usepackage[justification=centering]{caption}
\usepackage[Algorithm]{algorithm}
\usepackage{algpseudocode}
\pgfplotsset{compat=1.17}
\pgfplotscreateplotcyclelist{my cycle}{%
  color=blue!40!black\\% ,mark=*
  color=red!50!black, style=dashed\\% ,mark=+
  color=green!40!black, style=dotted\\% ,mark=diamond*
  color=orange!70!black,mark=triangle*\\%
  color=purple,mark=square*\\%
  color=cyan,mark=star\\%
  color=yellow,mark=otimes\\%
  color=red,mark=x\\%
  color=black,mark=pentagon*\\%
}

\definecolor{myblue}{HTML}{4f6fd6}

\tikzstyle{action} = [rectangle, rounded corners, 
minimum width=3cm, 
minimum height=1cm,
text centered, 
draw=black, 
inner sep=2mm,
text width=4cm,
fill=myblue,
text=white]

\tikzstyle{container} = [rectangle,
rounded corners, 
minimum width=3cm, 
minimum height=1cm, 
text centered, 
text width=3cm, 
draw=myblue,
thick]

\tikzstyle{label} = [rectangle,
rounded corners, 
minimum width=3cm, 
minimum height=1cm, 
text centered, 
draw=myblue,
text=myblue,
thick,
fill=lightgray!50]
\tikzstyle{arrow} = [thick,-{Latex[length=4mm]}]

\makeatletter
\g@addto@macro\@floatboxreset\centering
\makeatother

\newcommand{\RR}{\mathbb{R}}
\newcommand{\ZZ}{\mathbb{Z}}
\newcommand{\NN}{\mathbb{N}}
\DeclareMathSymbol{\shortminus}{\mathbin}{AMSa}{"39}
\newtheorem{remark}{Remark}

\usepackage{authblk}

\title{H-DES: a Quantum-Classical \\Hybrid Differential Equation Solver}
\date{\today}

\author[]{Hamza Jaffali}
\author[]{Jonas Bastos de Araujo}
\author[]{Nadia Milazzo}
\author[]{Marta Reina}
\author[]{Henri de Boutray}
\author[]{Karla Baumann}
\author[]{Frédéric Holweck}
\author[]{Youcef Mohdeb}
\author[]{Roland Katz}

\affil[]{ColibriTD, 91 Rue du Faubourg Saint Honoré, 75008 Paris, France\\
  \href{mailto:quantum@colibritd.com}{\textit{quantum@colibritd.com}}}
\begin{document}

\maketitle

\begin{abstract} 
In this article, we introduce an original hybrid quantum-classical algorithm
based on a variational quantum algorithm for solving systems of differential
equations. The algorithm relies on a spectral decomposition of the trial 
functions that are encoded directly in the quantum states generated by different 
parametrized circuits, and transforms the task of solving the
differential equations into an optimization problem. We first describe the
principle of the algorithm from a theoretical point of view. We provide a
detailed pseudo-code of the algorithm, on which we elaborate preliminary elements
for a complexity analysis to highlight some of its scaling properties. 
We apply our algorithm to a set of examples, running on emulators and real hardware 
showcasing its applicability across diverse sets of differential equations. We
discuss the advantages of our method and potential avenues for further
exploration and refinement.\footnote{The algorithm described in this paper is 
subject to a patent submission (EP24306601). Use of this method requires a 
license, which can be obtained from ColibriTD.}
\end{abstract}

% \begin{multicols}{2}

\section{Introduction}

The fine control of quantum systems is arguably among humanity's most remarkable
feats. The fragility of such systems, the core of the technological challenges
faced by researchers, provides both an advantage and a disadvantage. The former
motivates their use in quantum sensing beyond classical limits, as already
demonstrated using gravitational wave detectors~\cite{YMT+20} and portable
gravitometers with a sensitivity of over $10^{-9} g$~\cite{MVL+18}. The latter
hinders their use in simulations of quantum systems --- a prospect first
suggested by R. Feynman~\cite{Fey82}. Feynman's idea was to use controlled
quantum systems to simulate the behavior of natural systems. This task is
commonly challenging for classical computers, depending on the system's size and
symmetries.

~

The simulation of Hamiltonians is the foundation of analog quantum 
computing~\cite{Zak99}. A relevant problem is mapped onto the dynamics of a
quantum system, and the solution of the problem is usually encoded in the ground
state of a Hamiltonian. The task of finding this ground state is particularly
straightforward for quantum annealers, which apply the adiabatic 
theorem~\cite{BF28} to evolve an initial simple Hamiltonian into the final
Hamiltonian. In fact, quantum annealers using the Ising model~\cite{Gal99} have
already been manufactured and \textit{modest} claims of quantum advantage have
started to emerge~\cite{TAM+22}. This model can be translated into a quadratic
unconstrained binary optimization (QUBO) problem~\cite{KHG+14}, which finds
numerous applications. QUBO problems have a simple formulation, while their
optimal solutions, in contrast, are NP-hard to find by classical means. This
approach can be adapted to many combinatorial optimization problems in
finance~\cite{MKS+22}, vehicle routing~\cite{BGK+20}, logistics~\cite{SGM22}, 
isomer search procedures~\cite{TANM20}, and contact map overlaps~\cite{OSO18}, to 
name a few. Other controlled quantum systems can be implemented using Rydberg
atoms, which enable simulations of other dynamics, such as the XY Hamiltonian
\cite{HBS+20} and even ion traps, with more limited dynamics~\cite{ADG+22}.

~

Controlled quantum systems find a more versatile (often universal, in a
computational sense) use as quantum central processing units (QPUs). For this
purpose, many algorithms have been developed that provide speedups over their
best-performing classical counterparts, notably Shor's factoring and Grover's
search algorithms~\cites{Sho97,Gro97}. Recent studies have begun to unravel the
role played by superposition and entanglement in these speedups~\cites{JH19,
BJH+20}. These algorithms, however, demand high qubit numbers and
interconnectivity as well as long coherence times, much beyond what current
noisy intermediate-scale quantum (NISQ) devices can offer. A promising
alternative is provided by variational quantum algorithms (VQAs).

~

VQAs have the potential to perform well under NISQ-era limitations due to their
moderate hardware requirements, among which are shallow parametrized circuits of
$\lesssim 10^{2}$ qubits and the absence of all-to-all qubit gates~\cite{CAB+21}. 
The circuit parameters are optimized by a classical computer iteratively until
the generated quantum state satisfies pre-imposed conditions(encoded in loss
functions)~\cite{MBB+18}. Because of their potential for near-term applications,
their integration with high-performance computing (HPC), and their similarity to
machine learning techniques~\cite{WKS16}, many such algorithms have been
developed in several contexts. For example, the ground state of molecular
Hamiltonians can be found using a variational quantum eigensolver 
(VQE)~\cite{TCC+21} by means of the Jordan--Wigner transformation; a system of 
linear equations can be solved using the variational quantum linear 
solver~\cite{BLC+19} as opposed to the NISQ-nonviable HHL algorithm~\cite{HHL09}; 
a variational version of the quantum phase estimation (QPE) algorithm has been
developed in an effort to bring its usefulness into the NISQ era~\cite{FRF22}; 
and even combinatorial optimization problems are solvable via the quantum
approximate optimization algorithm (QAOA)~\cite{FGG14}.

~

Central to all these algorithms is the parametrized circuit structure, also
called the Ansatz, as its choice may enable faster convergence. In VQEs, for
instance, unitary coupled cluster Ansätze (plural of Ansatz)~\cite{PMS+14} are
outperformed in some metrics by adaptive derivative-assembled pseudo-Trotter
(ADAPT) ones~\cite{GEBM19}, which use fewer gates, reducing classical
optimization overheads. Nevertheless, the right choice of Ansatz, if it exists,
may not be sufficient to ensure convergence, since the classical optimization
may face barren plateaus, which prevent optimizers from finding the global
minima of objective functions~\cite{WFC+21}. These objective functions are
usually written in terms of expectation values, which need to be obtained
efficiently to avoid shot noise. In these cases, the naive measurement of
experimental outcomes can be replaced by approaches that require fewer shots,
such as classical shadows~\cite{HKP20}, its recent modifications~\cite{KG22}, and
Bayesian methods~\cites{MOB18,HH12,CG23}.

~

Quantum algorithms have been developed not only for tackling problems in which
classical computers perform poorly, but also for typically amenable problems,
such as solving differential equations (DEs). To name a few: the heat equation
was recently solved via a quantum annealer in a hybrid setup~\cite{PSG+21}; a
fully quantum method was used to solve the wave equation~\cite{CJO19}; and
another was used for the Navier--Stokes equations~\cite{Gai21}. Non-linear DEs
such as the latter are among the most challenging to solve via numerical methods
and have been the object of recently developed algorithms. In one instance, a
non-linear Schrödinger equation was solved by a VQA algorithm that included a
Hadamard test as a subroutine~\cite{LJM+20}. The latter approach is similar to
that of classical finite difference methods (FDM) as the function's domain is
discretized. The authors claim $\sim 20$ error-corrected qubits ($2^{20}$ grid
points) are enough to rival state-of-the-art supercomputers; the cost is
optimizing a number of circuit parameters that grows linearly with the number of
qubits. Another more versatile method resorts to differentiable quantum
circuits (DQCs)~\cites{MNKF18,SBG+19} and quantum feature maps~\cites{MNKF18,
LSI+20,SSM21} to evaluate derivatives without intrinsic numerical
errors~\cite{KPE21}. One possible drawback is that the number of $n$-qubit
circuits necessary grows as $\propto n^m$, where $m$ is the highest order of
derivatives in the DE.

~

In this paper, we propose an original variational quantum algorithm to solve partial
differential equations (PDEs) based on a spectral decomposition of the solution
function. The H-DES algorithm has been designed to run on NISQ devices in order to address industrial simulation demands with the available quantum machines.  The goal of the article is to present the general workflow of the algorithm, its flexibility and reliability, and how the related toolbox can be used in practice to solve DEs.

~

This article is organized
as follows. In Sec.~\ref{sec:description}, we introduce the algorithm and its
key principle and detail the steps of its workflow. In Sec.~\ref{sec:multivariate},
we discuss how the algorithm adapts 
to the case of multivariate equations. In
Sec.~\ref{sec:complexity}, we provide some premilinary theoretical elements concerning the
complexity and scaling of the hybrid algorithm. In Sec.~\ref{sec:results}, we run
the algorithm on several systems of DEs, showcasing the versatility of our
method. In Sec.~\ref{sec:discussions}, we discuss the advantages of our algorithm
and compare it to other approaches in the literature. In Sec.~\ref{sec:conclusion}, 
we conclude the article and present possible improvements of our algorithm. In the
appendices, we share how one can deal with boundary conditions in simple cases
(see App.~\ref{app:boundary}), some examples to illustrate the theoretical ideas
behind the algorithm (see App.~\ref{app:examples}), more details from the
complexity study (see App.~\ref{app:runtimeDetails}), and a theoretical framework
for performance comparisons of several solvers (see 
App.~\ref{app:performance_profile}).

\section{Description of the algorithm}
\label{sec:description}

In this section, we present the overall principle and detail the main steps of the
algorithm. We explain the theoretical basis on which our algorithm is built, for
which pedagogical examples can be found in App.~\ref{app:examples}. For
simplicity, we first discuss our algorithm in the context of ordinary DEs
(ODEs) and then show in Sec.~\ref{sec:multivariate} how the method can naturally
be extended to solve partial DEs (PDEs).

~

\begin{figure}[H]
  \centering
  \input{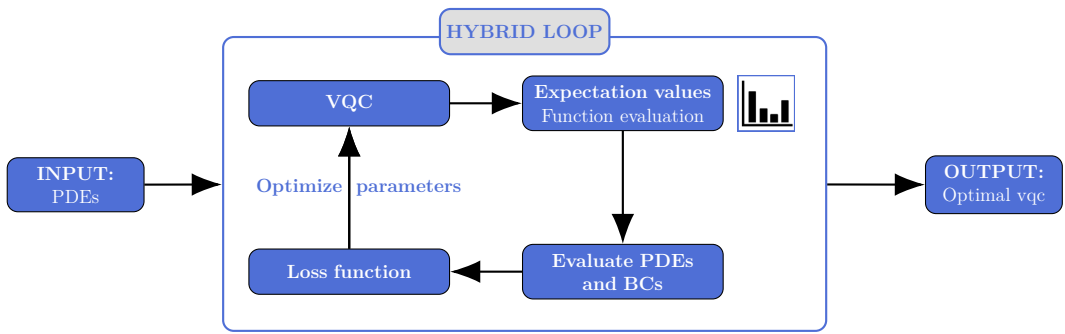}
  \caption{An illustration of the workflow of our approach.}
  \label{fig:workflow}
\end{figure}

We start with an overview of the hybrid quantum-classical algorithm for solving
DEs. We illustrate the general workflow of our algorithm in 
Fig.~\ref{fig:workflow} and list the general steps below:

\begin{enumerate}
  \item We take as input the PDEs and their boundary conditions;
  \item We encode the problem in a loss function that gives us a measure of how 
    far the current attempt is from the solution;
  \item We generate trial states from parametrized VQCs;
  \item We measure expectation values of observables with respect to these
    states to evaluate trial functions and their derivatives;
  \item We compute the loss function using these evaluations;
  \item Using a classical optimizer, we update the parameters of the circuits 
    to minimize the loss function;
  \item The hybrid loop, consisting of the four previous steps, is repeated 
    until a given error tolerance is reached. The optimal parameters are then
    returned and the solution functions can be retrieved.
\end{enumerate}

We further detail the steps of the algorithm in the following subsections and
provide a pseudo-code in Sec.~\ref{sec:pseudo_code}.

\subsection{Variational Quantum Circuit and Ansatz}

To construct the parametrized state used to minimize the loss function, we need a
variational quantum circuit (VQC). The architectures employed here 
are usually based on the structure of the hardware-efficient Ansatz (HEA), but not limited to it. In typical HEA setup, each set of parametrized single-qubit rotations is followed by an entangling layer of controlled (usually CNOT) gates. This block of rotations and controlled gates is then repeated $d$ times, where $d$ is called the depth of the VQC. In what follows, we consider linear connectivity among qubits, but the controlled gates can be easily generalized to other connectivity architectures~\cite{LOCC24}.

~

As $d$ increases, the number of rotation parameters $\theta$ increases
proportionally, giving rise to a circuit with higher expressivity power. These
layers make the VQC structure comparable to those of classical neural networks,
with rotation gates playing the role of neurons and angles playing the role of
synaptic weights. By tuning and adjusting the circuit's parameters, we change
the output state to minimize the loss function.

~

\begin{figure}[H]
  \centering
  \input{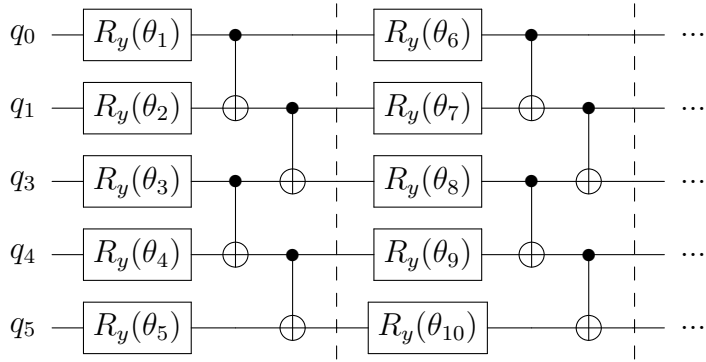}
  \caption{A representation of the hardware-efficient Ansatz used in our 
    variational quantum algorithm.}
  \label{fig:ansatz}
\end{figure}

Generally, the rotation layer includes rotations around different axes, i.e.,
$R_x$, $R_y$, and $R_z$ gates, but in the case of a basis-state encoding (see Sec. \ref{sec:encoding}), 
a possible choice for this layer is to only use
parametrized $R_y$ gates for each qubit (see Fig.~\ref{fig:ansatz}). This choice
is driven by the fact that, in the case of basis-state encoding, we do not need to generate states with complex 
amplitudes since only the associated probabilities are used to encode the 
solution function. We thus chose to restrict the research space to real states, 
for which using only $R_y$ gates is sufficient. We also empirically observed 
that results for different loss functions reached satisfactory precision 
with a single rotation gate per layer (at least in the ideal case, i.e., with 
perfect operations in the circuit).

When running on real QPUs (Quantum Processing Unit), the circuit has to go through a transpilation process to express the circuit by means of unitary gates supported by the QPU (also called native gates). In that case, building the ansatz directly by using native gates can help to avoid extra-transpilation steps that usually increase the depth of the circuit and thus the risk of gate-noise propagation and decoherence during execution. Taking into account the qubit connectivity directly in the selection of qubits or the design of the ansatz also helps to reduce the number of additional gates (and thus noise) introduced at the transpilation step.

\subsection{Spectral decomposition and Chebyshev polynomials}

Our approach is based on the spectral decomposition, in which the solution function is
expressed as a linear combination of the elements of a complete basis of
orthogonal functions. Therefore, the task of finding the best solution to a
DE consists of determining the best approximation of the
coefficients of the solution in that basis.

~

One possible choice is to express the solution function in the basis of Chebyshev polynomials,
which are widely used as basis functions for spectral methods~\cites{Boy01,WS07}.
These polynomials form a complete orthogonal system and allow us to represent
piecewise smooth and continuous functions in finite domains. The expression of
their derivatives is straightforward and can also be defined by recurrence,
making it easy and efficient to implement.

~

One can find different definitions of Chebyshev polynomials in the literature. In
our algorithm, we use the ``first kind'', defined as follows:

\begin{equation}
  \text{Cheb}(k, x) = \\
  \begin{cases}
    \cos(k\arccos x)                           & \text{ if } |x| \leq 1 \\
    \cosh(k\operatorname{arcosh} x)            & \text{ if } x \geq 1   \\
    (-1)^{k}\cosh(k\operatorname{arcosh} (-x)) & \text{ if } x \leq -1
  \end{cases} ~,
\end{equation}
with $x\in\RR$ the point at which we evaluate the polynomial and $k \in \ZZ^{+}$
the order of the Chebyshev polynomial.

~

Therefore, the expression of a given function $f: \RR \to \RR$ expressed in the
truncated Chebyshev polynomial basis, with $C$ terms, can be written as
follows:

\begin{equation}
  f(x) = \sum_{k=0}^{C-1} c_k ~\text{Cheb}(k, x) ~,
\end{equation}
with $c_0, c_1, \dots, c_{C-1} \in \RR$.

~

In the H-DES algorithm, the choice of the basis is not only limited to Chebyshev polynomials, but allows one to select any other suitable complete basis (Legendre, Laguerre, Fourier, B-Splines and NURBS, Walsh functions, ...), or even modify and combine several bases. This is a possible way to impose exactly some boundary conditions. In the following sections, we will, by default, discuss the principle of the algorithm by considering Chebyshev polynomials as a basis choice.

\subsection{Encoding and evaluation of the function}\label{sec:encoding}

In this subsection, we detail how we use the $n$-qubit state generated by the
variational quantum circuit to represent the solution function. We denote as
$\ket{\psi_f}$ the state coming out of the VQC modeled by the gate $U_{\theta}$, 
expressed in the computational basis (with decimal notation) as follows:

\begin{equation}
\ket{\psi_f} = U_{\theta}\ket{0}^{\otimes n} = \sum_{i=0}^{2^n -1} a_i \ket{i} ~,
\end{equation}

\begin{equation}
\text{with } ~~ \sum_{i=0}^{2^n -1} |a_i|^2 = 1 ~.
\end{equation}

\subsubsection{Basis-state encoding}

We encode the solution function $f$ in the amplitudes of the state, or more
precisely, in the associated probabilities $p_i$. We associate a basis state
with each Chebyshev polynomial (of a certain order), and the coefficient in
front of this polynomial becomes the probability of measuring that basis state.

~

Since probabilities are positive real numbers, expressing $f$ using only positive
coefficients in the basis would not allow us to represent all continuous
functions. One would also need to combine positive and negative coefficients in
the decomposition of the solution function. In order to do this, we add to the
list of polynomial values a copy of each one, multiplied by $-1$. In the quantum 
encoding of our function, these probabilities are associated with the second 
half of the basis states (see Eq.~\ref{eq:obs-cheb}).

~

Furthermore, in the particular case $x\in[-1,1]$, we have $\text{Cheb}(k, x) \in
[-1,1]$ for all $x$ and $k$. When we take the linear combination of Chebyshev
polynomials with the associated difference of probabilities (because of the
duplication of Chebyshev polynomials with a minus sign), which are bounded by
the normalization constraint, the resulting function is also bounded. Therefore,
to allow for the modeling of any general solution function, we introduce a
scaling parameter $\lambda \in \RR$.

~

Finally, we encode the solution function $f$ in the following way:

\begin{equation}\label{eq:expression_solution}
f(x) = \lambda\sum_{i=0}^{2^{n-1}-1} (p_i - p_{i+2^{n-1}})~\text{Cheb}(i, x)~,
\end{equation}
with $p_i = |a_i|^2$.

~

Therefore, the variational circuit, for a given set of parameters 
$\theta=\{\theta_i\}_i$, models the state that encodes the decomposition of the 
function in the Chebyshev polynomial basis. Note that in the case of a system of
DEs, each solution function has a different circuit, and one can choose the number
of qubits to allocate for each circuit independently.

~

To obtain or evaluate the solution function, one needs to retrieve the properties
of its state. Since we are only concerned with the probabilities, there is no
need to perform a full state tomography, since the probabilities correspond to
the diagonal terms of the density matrix describing the state. Nevertheless, we
still have to retrieve $N=2^n$ terms, which becomes difficult for $n\gg1$
\cites{MJZ+16,FB18, GXLK21,STN+14}.

~

Instead, we determine the value of the solution $f$ at a point $x$,
following the definition in Eq.~\ref{eq:expression_solution}, by computing the
expectation value with respect to a diagonal observable $O_C(x)$ such that:

\begin{equation}
f(x) = \lambda ~ \expval{O_C(x)}{\psi_f}~,
\end{equation}
with

\begin{equation}\label{eq:obs-cheb}
\begin{gathered}
O_C(x) = \sigma_z \otimes \left(\sum_{i=0}^{2^{n-1}-1} \text{Cheb}(i,x) \ketbra{i}{i} \right) \\
       = \begin{pmatrix}
  \text{Cheb}(0,x)                                                                                              \\
                   & \ddots &                          &                   &        & \text{\huge0}             \\
                   &        & \text{Cheb}(2^{n-1}-1,x)                                                          \\
                   &        &                          & -\text{Cheb}(0,x)                                      \\
  \text{\huge0}    &        &                          &                   & \ddots                             \\
                   &        &                          &                   &        & -\text{Cheb}(2^{n-1}-1,x) \\
 \end{pmatrix} ~.
\end{gathered}
\end{equation}

~

By computing this expectation value, one can evaluate the function at a given
point. This does not allow us to retrieve the full expression of the function
$f$, but in the context of our VQA, we only need, at each iteration, to evaluate
the function at specific points (see Sec.~\ref{sub:loss}). We present a simple
example to illustrate the encoding and evaluation of the functions in
App.~\ref{sec:modeling}.

~

In the following sections, we will, by default, discuss the principle of the algorithm by considering the global and diagonal observables corresponding to basis-state encoding. The number of probabilities to be estimated grows exponentially with the number of qubits. Moreover, the global nature of the observable implies that the cost function is also global, which may lead to the appearance of barren plateaus, which is another significant limitation on the range of applicability. In the next subsection, we introduce an alternative encoding to overcome these issues.

\subsubsection{Pauli-monomial encoding}

Another way of encoding the spectral basis in the observable is to consider a linear combination of Pauli monomials instead of probabilities of basis states. This provides another general framework to compute the evaluation of the trial function, with more freedom on the number of orthogonal functions and the choice of the Pauli monomial appearing in the decomposition, which will determine the size of the spectral basis. This is particularly interesting when dealing with barren plateaus, measurement strategies, and readout error \cites{cerezo2021, kashif2025}.

~

Therefore, in the case of a Pauli monomial encoding, the observable $O_C(x)$ can be expressed as the linear combination of $M \in \NN^*$ Pauli monomials such that:

\begin{equation}
O_C (x) = \sum_{i=0}^{M-1} o_i \cdot \text{Cheb}(i,x) \cdot P_i ~,
\end{equation}

with, for all $i \in \llbracket 0, M-1 \rrbracket$, $o_i \in \RR^*$ a ponderation factor and $P_i \in \{I,X,Y,Z\}^{\otimes n}$ an element of the set of generalized Pauli operators, also called $n$-qubit Pauli monomial.

~

A specific subset of the generalized Pauli ensemble of interest can be chosen to define the loss function more locally. Indeed,  the $k$-local observables \cites{kitaev2002, peruzzo2014,cubitt2016,cerezo2021} are defined as observables for which the Pauli monomials appearing in their decomposition are acting non-trivially on at most $k$ qubits. A given Pauli monomial $P_i = P_{i,1} \otimes P_{i,2}  \otimes  \cdots \otimes P_{i,n} $, is said to act non-trivally on $k$ qubits (also called $k$-body interaction) if there are exactly $n-k$ Pauli atoms $P_{i,j}$ that are the (one-qubit) identity operator.

~

The maximum possible number $M_k$ of Pauli monomials appearing in the decomposition of a general $k$-local observable is given by:

\begin{equation}
M_k = \sum_{i=1}^{k} 3^i \binom{n}{i} ~.
\end{equation}

\subsection{Evaluation of the derivatives}

In the case where we chose Chebyshev polynomials as a basis, one of the most remarkable advantages of this algorithm is the simplicity with which any derivative of a function can be calculated, and this is due to two characteristics of this method. The first characteristic is that a derivative of any order can
expressed as
\begin{equation}
\frac{\partial^q f(x)}{\partial x^q} = \lambda \sum_{i=0}^{2^{n-1}-1}
  (p_i - p_{i+2^{n-1}})~\frac{\partial^q \text{Cheb}(i, x)}{\partial x^q} 
\end{equation}
and calculated by means of the same circuit as an expectation value of
the observable
\begin{equation}
\frac{\partial^q f(x)}{\partial x^q}=\lambda~\expval{O_{\partial^q C}(x)}{\psi_f}~,
\end{equation}
where $O_{\partial^q C}(x)$ has exactly the same structure as $O_{C}(x)$ but
with all the diagonal elements replaced by the $q$-order derivative of the
corresponding Chebyshev polynomial. The second favorable characteristic is
actually related to the Chebyshev polynomials directly, since any derivative
${\partial^q\text{Cheb}(i, x)}/{\partial x^q}$ can be defined using
the following expression:
\begin{equation}
\frac{\partial^q \text{Cheb}(i, x)}{\partial x^q} = 
  2^q i \sideset{}{'}\sum_{\substack{0\le k \le i-q\\k\equiv i-q~(\text{mod }2)}}
    \binom{\frac{i+q-k}{2}-1}{\frac{i-q-k}{2}}
    \dfrac{(\frac{i+q+k}{2}-1)!}{(\frac{i-q+k}{2})!}
    \text{Cheb}(k, x),
\end{equation}
where a prime after a summation symbol means that the term contributed by $k=0$
is to be halved, if it appears.

\subsection{Loss function}
\label{sub:loss}

The loss function $L$ encodes in itself the DEs we want to solve, giving us a way
to quantify the difference between the solution of the problem considered and the
trial functions we reconstruct by means of the observables. Using this value, the
classical optimizer can update the parameters of the circuit with the ultimate
aim of reducing this distance. 

~

In order to build the loss function, we collect all the terms of the DE on the
left side, i.e., $E_i(x) = 0$, where $E$ is the set of differential equation in
our system, and we thus consider the $i^{th}$ differential equation. The 
evaluation of the DEs requires the evaluation of all the functions and their 
derivatives (appearing at least once) at each sample point, (sometimes called collocation points), which in turn 
implies the computation of expectation values of the right observables.

\begin{remark}
\label{remark:linear}
In the case of a linear DE involving only one function, one can evaluate it for
a specific point $x_s$ by evaluating only one expectation value. In fact, the DE
is defined as the linear combination of the solution function, its derivatives,
and potentially the sample point itself. These terms are evaluated by computing
the expectation value of an observable ($O_{\partial^q C}(x)$ for its $q^{th}$ 
derivative, and the scaled identity $x_s  I$ for the sample point) on the same 
state. Therefore, the DE can be evaluated by regrouping all the observables into 
a single one using the linearity of the inner product, which will make the algorithm essentially similar to a Variational Quantum Eigensolver. This is also possible for non-linear DEs, when the Pauli monomials in the decomposition of the observables (for function and derivatives evaluation) are all commuting. In that case, using a Pauli grouping process, it is sufficient to evaluate the expectation value of a single monomial to retrieve the others by post-processing.
\end{remark}

~

The solution of a (system of) PDE(s) must respect both the functional and boundary
conditions. These two criteria can be taken into account in the loss function as

\begin{equation}
  L(\theta) = L^{\rm{diff}}(\theta) + \eta \cdot L^{\rm{boundaries}}(\theta) ~,
\end{equation}
where the first term evaluates the residual of the DEs, while the 
second accounts for the boundary conditions. 

~

The coefficient $\eta$ controls the weight of the boundary terms in the
optimization and has to be chosen carefully to ensure that the constraint is
imposed sufficiently. To tackle this task, called \textit{loss balancing} in the
context of multi-objective optimization, several strategies have been proposed 
in the deep-learning literature~\cites{BK21,HTM19,WTP21}. In some cases, it 
can be chosen so it is higher than the maximal possible values of the left term,
and this represents a well-known problem in constrained optimization~\cites{WS07,
NW06}. Adding a term directly to the loss function allows us to incorporate
simple and complex boundary conditions within the same framework. When the
boundary condition is simple or the functional form of the solution is
reasonably independent of it, one can consider other ways of handling the
boundary conditions (see App.~\ref{app:boundary}), and in that case we set
$\eta=0$.

~

To construct both terms, one can use the mean squared error (MSE). Using MSE metrics,
the differential part of the loss function can be explicitly written as

\begin{equation}
L^{\rm{diff}}(\theta) = 
  \frac{1}{n_s}\sum_{e \in E}\sum_{x_s \in S} e(x_s)^2 ~,
\end{equation}
where the $x_s$ are the coordinates of the sample points, $\theta$ is the set of
angles parametrizing the circuit. Similarly, the boundaries component can be computed
as the sum of all errors for each boundary condition for all functions appearing
in the DE. This can be expressed as follows:

\begin{equation}
L^{\rm{boundaries}}(\theta) =
  \frac{1}{n_{BC}}\sum_{f \in F}~\sum_{f_{BC}, x_{BC} \in BC(f)}
    \big(f(x_{BC}, \theta) - f_{BC}(x_{BC})\big)^2 ~,
\end{equation}
with $x_{BC}$ and $f_{BC}$ the point and function, respectively, representing the
boundary conditions for each function $f\in F$ and $n_{BC}$ the total number of
boundary conditions.

\subsection{Validation of the results}
\label{sec:validation}

After retrieving the solution proposed by our hybrid solver, we want to evaluate
the quality of the result. In this case, instead of replacing the solution
functions in the DEs, as we do in the algorithm loop, we compare the solution to
that provided by a classical solver (assuming that the latter returns the exact
expression of the solution).

~

For each function $f \in F$ involved in the DEs to be solved, the expected
solution $c_f$ is computed. The idea is to evaluate the quality of each solution
function $f$ by computing some distance to $c_f$. We consider two definitions of
the distance: $d_1$ and $d_2$. In both definitions, we compare the evaluation of
the functions for a set of samples $S_v$, ideally containing many more elements
than the set $S$ used for the optimization.

~

The first distance $d_1$ is computed by taking the maximum over $S_v$ of the
absolute difference between the evaluation of the two functions:

\begin{equation}
d_1(f,g) = \max_{x \in S_v} |f(x) - g(x)| ~.
\end{equation}

The second distance is the average of the squared difference between the
evaluation of the two functions throughout $S_v$. Using the average makes the
comparisons of $d_2$s computed from differently sized $S_v$s more relevant.

\begin{equation}
  d_2(f,g) = \frac{1}{|S_v|} \sum_{x \in S_v} \big( f(x) - g(x) \big)^2 ~.
\end{equation}

Both distances provide different information about the quality of the fit. The
first distance provides us with information about the maximum error of the
solution function, which is useful if one needs a perfect fit at each sample
point. The second distance indicates whether the solution overall is close to
the true value, with potentially large or small variances in the errors.
Therefore, by combining both distances, we can retrieve comprehensive
information about the overall error as well as partial information on the
variance. Let $V_{f,0}$ be the validation score for the function $f$, defined as
the following tuple:

\begin{equation}
  V_{f,0} = \Big(d_1(f,c_f) , ~ d_2(f,c_f) \Big) ~.
\end{equation}

However, fitting the function on the sample points does not necessarily ensure
that the solution function, expressed in the spectral basis, satisfies the DE.
One also has to ensure that the derivatives of the solution function
(those appearing in the DE at least) match the derivatives of the actual
solution. We thus extend the validation score $V_{f,0}$ to incorporate higher
derivatives of the solution function as follows: 

\begin{equation}\label{eq:valid_deriv}
  V_{f,i} = \Big(d_1(f^{(i)},{c_f}^{(i)}) , ~ d_2(f^{(i)},{c_f}^{(i)}) \Big) ~,
\end{equation}
where $f^{(i)}$ is the $i^{th}$ derivative of the function $f$. One can then
combine the validation scores of the function $f$ and its derivatives to define
another overall score for the quality of the solution $f$. We denote as 
$G_F(f)$ the set of orders of derivatives of the function $f$ appearing in the
set of DEs to solve and define the validation score $V_f$ such that:

\begin{equation}
V_f = \Big(
  \max_{i \in G_F(f)} ~ V_{f,i}(0), ~ 
  \frac{1}{|G_F(f)|} \sum_{i \in G_F(f)} V_{f,i}(1)
\Big) ~,
\end{equation}
with $V_{f,i}(a)$ referring to the $a^{th}$ ($a \in \{0,1\}$) entry of the validation
couple defined as $V_{f,i}$ in Eq.~\ref{eq:valid_deriv}. This validation score
$V_f$ gives us information on the quality of each solution function $f$ and
can be compared with other validation scores since it is only defined using
absolute differences, and their range of values is also independent of the
number of samples in $S_v$.

~

We finally define a global validation score $V$ of the solver by combining each
function's validation score $V_f$, such that:

\begin{equation}
V = \Big(\max_{f \in F} ~ V_f(0), ~ \frac{1}{|F|} \sum_{f \in F} V_f(1) \Big) ~.
\end{equation}

The first value gives us information on the maximum error among all functions and
sample points. Having a high first value warns us that for some points, the
solution function can be very far from the expected result. The second value
gives us information on the overall error among all functions and sample points.
Having a low second value ensures that overall, the solution functions live
around the true solution. If one value of the validation couple is not
satisfying, one can compute the intermediate validation score $V_f$ to determine
which solution function of the system of DEs is responsible for the error in the
validation. If one wants to identify exactly whether the problem comes from a
solution function or its derivatives, it is still possible to compute all the $V_
{f,i}$ scores. Otherwise, if the global validation score satisfies some criteria
or thresholds, it is sufficient to conclude that all the solutions and their
derivatives are valid. 

~

We thus obtain a validation score that can be used to compare solutions for
different sets of DEs, different solver parameters, or different solvers
entirely. The genericity of the validation score allows us to then establish an
acceptance threshold below which we consider the set of DEs solved. This can be
dictated by the precision needed in different fields or industries, or, for
instance, by benchmarking several sets of DEs and tuning the threshold to match a
given acceptance criterion. In App.~\ref{app:performance_profile}, we present
a theoretical framework to describe and compare the quality of convergence of
several solvers, within which the global validation score $V$ can be used to
measure performance.

\section{Multivariate functions}
\label{sec:multivariate}

In this subsection, we detail the multivariate case, i.e., when a solution
function $f$ takes as input $X=(x_1,x_2,\dots,x_v) \in \RR^v$. Since the
variable dependence is encoded only in the observables, moving from
1-dimensional to $v$-dimensional space does not affect the rest of the
algorithm. We recall that we focus on the basis-state 
encoding of the trial function, in the basis of Chebyshev polynomials.

~

In the unidimensional case, each basis state (of the computational basis) is
used to represent, in binary notation, the order of a Chebyshev polynomial.
Except for the first qubit, which is used to determine the sign in front of the
polynomial, all the remaining qubits of the basis states are indeed used to
encode the corresponding order. In the multivariate case, we assign to each
variable $x_j$ a specific number of qubits that are used to encode the order
of the corresponding Chebyshev polynomial. In fact, one possible way of defining
the multivariate Chebyshev polynomial $\text{Cheb}(i, x_1,x_2,\dots,x_v)$ is to
take the product of independent univariate Chebyshev polynomials $\text{Cheb}
(L_j,x_j)$ defined on a rectangular domain~\cites{PV11,Tre16,Mas80}.

~

Therefore, each basis state $\ket{i}$ of the $(n+1)$-qubit state modeling the
solution function is used to encode the different indices assigned to each variable.
We write the binary decomposition of the basis state $\ket{i}$ (written in
decimal notation) as
\begin{equation}
\ket{i} = \ket{i_0}\ket{i_1i_2i_3\dots i_{n}} ~.
\end{equation}

For each variable $x_j$, we attribute $l_j$ qubits to encode the order (of the
Chebyshev polynomial) and denote as $L_j$ the binary decomposition of the order.
This allows us to separate the basis state $\ket{i}$ as the tensor product
of the first (sign) qubit with $v$ states, such that

$$\ket{i} = \ket{i_0}\ket{i_1\dots i_{l_1}}\ket{i_{l_1+1} \dots
    i_{l_1+l_2}}\cdots\ket{i_{l_{v-1}+1}\dots i_{n}} =
  \ket{i_0}\ket{L_1}\ket{L_2}\cdots\ket{L_v} ~,$$
with
$$l_1 + l_2 + \cdots + l_v = n ~.$$

In this way, we allow freedom in the selection of the required number of qubits
$l_j$ for each variable $x_j$, which can be useful when some variables require
more terms in the spectral decomposition. To summarize, we evaluate the
multivariate Chebyshev polynomial using the following formula:

$$\text{Cheb}(i, X) = \text{Cheb}(i, x_1,x_2,\dots,x_v) = \prod_{j=1}^{v}
  \text{Cheb}(L_j,x_j) ~.$$

This allows us to write the corresponding encoding of a multivariate solution
function $f$ with the same notation as before:

$$f(x_1,x_2,\dots,x_v) = \lambda \sum_{i=0}^{2^{n-1}-1} (p_i -
  p_{i+2^{n-1}})~\text{Cheb}(i, x_1,x_2,\dots,x_v) ~.$$

Using the same idea, the only change appearing when evaluating the function at a
sample point is in the expression of the observable $O_C$:

$$f(x_1,x_2,\dots,x_v) = \lambda ~ \expval{O_C(x_1,x_2,\dots,x_v)}{\psi_f}~,$$
with

$$O_C(x_1,x_2,\dots,x_v) = \sigma_z \otimes \left( \sum_{i=0}^{2^{n-1}-1}
  \text{Cheb}(i,x_1,x_2,\dots,x_v) \ketbra{i}{i} \right) ~.$$

We present in App.~\ref{sec:multivariate_example} a simple example to
illustrate the encoding principle in the multivariate case.

\subsection{Computing the derivatives efficiently}

When we move to the multivariate case, the computation of the partial derivatives
of the solution function can become more complex. However, since the
multivariate Chebyshev polynomial is expressed as a product of independent
single variable Chebyshev polynomials, it makes the derivatives easier to
express. For instance, in the case of a partial derivative of order $g = 1$, we
can write:

\begin{equation}
  \frac{\partial \text{Cheb}}{\partial x_j}(i, x_1,x_2,\dots,x_v)
  = \frac{\partial \text{Cheb}}{\partial x_j}(L_j , x_j)
  \times \prod_{k=1, k\neq j}^{v} \text{Cheb}(L_k,x_k) ~.
\end{equation}

We point out that for computing the partial derivative with respect to the $j^{th}$
variable, we only need to compute one partial derivative. Furthermore, the
rightmost term of the product corresponds to the product of all the remaining
univariate Chebyshev polynomials that are already computed to evaluate the
solution function. This large product is also equal to the value of the
$v$-dimensional Chebyshev polynomial evaluated at the sample point of $\RR^v$
divided by the univariate Chebyshev polynomial evaluated on the $j^{th}$ coordinate
of the sample point. In other terms, we have

\begin{equation}
  \frac{\partial \text{Cheb}}{\partial x_j}(i, x_1,x_2,\dots,x_v)
  = \frac{\partial \text{Cheb}}{\partial x_j}(L_j , x_j)
  \times \frac{\text{Cheb}(i,x_1,x_2,\dots,x_v)}{\text{Cheb}(L_j , x_j)} ~,
\end{equation}
which allows to reduce the number of multiplications needed and reuse already
computed evaluations of the Chebyshev polynomials (uni- and multi-variate).
Therefore, by computing all first derivatives of the univariate Chebyshev
polynomials, we can compute any first-order derivative of the multivariate
Chebyshev polynomial.

~

When we move to the case of second-order derivatives ($g=2$), we need to
distinguish between two cases: that of a partial derivative that is second order
in a single variable and that of a partial derivative in two distinct variables.
In the first case, we only need to elevate the derivative to the second order as
follows:

\begin{equation}
\frac{\partial^2 \text{Cheb}}{\partial {x_j}^2}(i, x_1,x_2,\dots,x_v) = 
  \frac{\partial^2 \text{Cheb}}{\partial {x_j}^2}(L_j , x_j) 
    \times \frac{\text{Cheb}(i,x_1,x_2,\dots,x_v)}{\text{Cheb}(L_j , x_j)} ~.
\end{equation}

Since the second case is a product of functions of independent variables,
we retrieve the product of both derivatives with the product of the remaining
Chebyshev polynomials. By recalling that $X=(x_1,x_2,\dots,x_v)$, we obtain:

\begin{equation}
\frac{\partial^2 \text{Cheb}}{\partial x_{j_1}\partial x_{j_2}}(i, X) = 
  \frac{\partial \text{Cheb}}{\partial x_{j_1}}(L_{j_1} , x_{j_1}) 
    \times \frac{\partial \text{Cheb}}{\partial x_{j_2}}(L_{j_2} , x_{j_2}) 
    \times \frac{\text{Cheb}(i,X)}{\text{Cheb}(L_{j_1} , x_{j_1}) 
    \times \text{Cheb}(L_{j_2} , x_{j_2})} ~.
\end{equation}

This result can be generalized to any partial derivative of order $g$ with
respect to the variables denoted by $x_{j_1}, x_{j_2}, \dots, x_{j_g}$. Some of
these variables can appear several times in the partial derivative. We suppose
that these $g$ variables can be regrouped into a set of $h$ $(\leq g)$ distinct
variables $x_{k_1}, x_{k_2}, \dots, x_{k_h}$. We denote as $g_l$ the number of
times the variable $x_{k_l}$ appears in the partial derivative, and we thus have
$g_1 + g_2 + \cdots + g_h = g$. Therefore, we can rewrite the expression for the
$g^{th}$ derivative in the following manner:

\begin{equation}
\frac{\partial^g \text{Cheb}}{\partial x_{j_1}\dots\partial x_{j_g}}(i, X) = 
  \frac{\partial^g \text{Cheb}}{\partial^{g_1} x_{k_1}\dots\partial^{g_h} x_{k_h}}(i, X) ~,
\end{equation}
such that:

\begin{equation}\label{eq:g_leq_half}
\frac{\partial^g \text{Cheb}}{\partial^{g_1} x_{k_1}\dots\partial^{g_h} x_{k_h}}(i, X) = 
  \prod_{l=1}^{h} \frac{\partial^{g_l} \text{Cheb}}{\partial^{g_l} x_{k_l}}(L_{k_l} , x_{k_l}) 
    \times \frac{\text{Cheb}(i,X)}{\prod_{l=1}^{h} \text{Cheb}(L_{k_l} , x_{k_l})} ~.
\end{equation}

In the case where $h$ is greater than half the number of variables, it is more
efficient, for the second term of the product, to directly compute the product
of the remaining variables instead of dividing the multivariate Chebyshev
polynomial, which leads us to:

\begin{equation}\label{eq:g_geq_half}
\frac{\partial^g \text{Cheb}}{\partial^{g_1} x_{k_1}\dots\partial^{g_h} x_{k_h}}(i, X) = 
  \prod_{l=1}^{h} \frac{\partial^{g_l} \text{Cheb}}{\partial^{g_l} x_{k_l}}(L_{k_l} , x_{k_l}) \times \prod_{j\notin \{k_l\}_l}  \text{Cheb}(L_{j} , x_{j}) ~.
\end{equation}

For the computation of the $g^{th}$ derivative, we need to perform 2$h + 1$ 
multiplications and one division, following Eq.~\ref{eq:g_leq_half} (or $n_v + 1$
multiplications if $h \geq \frac{n_v}{2}$ according to Eq.~\ref{eq:g_geq_half} ).

~

In the worst case, we can imagine that all the possible partial derivatives are
involved in the DEs and that we will have to compute them all. In this scenario
(supposing that we already computed all the Chebyshev functions needed), we must
compute

\begin{equation}
  \sum_{g=1}^{n_d} \binom{n_v + g-1}{n_v}
\end{equation}
partial derivatives, which correspond to the sum of all possible partial
derivatives at each order, with $n_d= \max(G_F)$ the highest derivative order
involved in the DEs, and $n_v$ the maximum number of variables in functions
involved in the DEs (see next section for the definition of $G_F$). Consequently, 
the computational cost of computing all the partial derivatives, expressed as 
a function of the cost $C_b$ of a multiplication/division, is equal to:

\begin{equation}
\sum_{g=1}^{n_d} \binom{n_v + g-1}{n_v} \times (2h_d +1) \times C_b = 
  \frac{n_d \times \binom{n_d + n_v}{n_v}}{n_v+1} \times (2h_d +1) \times C_b ~.
\end{equation}

To compute all these partial derivatives, we only need to evaluate
and multiply the value of the $d^{th}$ partial derivatives in one variable 
$\frac{\partial^d\text{Cheb}}{\partial{x_j}^d}(L_j, x_j)$. For each value of $d$, 
we have $n_v$ such partial derivatives to compute. In the worst case, we have to
prepare all the possible partial derivatives of that form. This means that, for
half of the basis states (since the other half only introduces a minus sign in
front of the Chebyshev polynomials) and all sample points, in the worst case, we
compute and store $2^{n-1} \times n_v \times n_d \times n_s$ values.

\section{Implementation and related complexity}
\label{sec:complexity}

In this section, we provide preliminary elements for a theoretical study 
of our algorithm to examine the
complexity of each step. More precisely, we aim to use theoretical arguments to
study the algorithm scaling properties (for both the quantum and classical parts)
as a function of the inputs of the problem to be solved (e.g., the number of
sample points, number of variables, and number of DEs). The goal of this
analysis is to provide initial insights on the stability and realism of our method 
when moving to more complex systems of DEs. In the following, we will mainly 
focus on the estimation of the computational time.

\subsection{Pseudo-code of the algorithm}\label{sec:pseudo_code}

In this subsection, we present a pseudo-code (\ref{alg:hdes}) of the algorithm listing its main steps. The algorithm takes as input all information needed to define the differential equations to solve, as well as the required number of qubits $n$, the depth of the circuit $d$, the set of sample points $S$, the number of sample points $|S|=n_s$, and the target precision $\varepsilon$ for the loss function. The set $F$ contains the set of all functions that appear in the set of differential equations $E$. We denote as $G_F$ the set of all orders of derivatives (using an order of 0 for the function itself) appearing in $E$ for each function in $F$. We denote as p the parameters of the VQC that are inputs of the \textit{EVAL} function.

\begin{algorithm}[H]
\caption{H-DES (Hybrid Differential Equation Solver)} \label{alg:hdes}
\begin{algorithmic}[1]
\Require $E$ : the set of differential equations to solve
\Require $F$ : the set of functions involved in $E$
\Require $G_F$ : the set of orders of derivatives involved in $E$ for each function in $F$
\Require $D$ : the set of domains of definition for each function in $F$
\Require $n$ : the number of qubits
\Require $d$ : the depth of the VQC
\Require $n_s$ : the number of sample points
\Require $\varepsilon$ : the target precision for the error
\Require $m_{iter}$ : the maximum number of iterations of the optimizer
\Ensure Approximation of every function $f \in F$ \vspace{.5cm}

\State listObs, S $\gets$ \Call{GenerateObservables}{$G_F, D, n, n_s$}\label{line:obs}
\State qc $\gets$ VQC($n$, $d$) \label{line:vqc}
\vspace{.3cm}
\Function{Eval}{p}
  \State e$_{vals} \gets$ exp\_value(qc(p), listObs) \label{line:expval}
  \State \textbf{return} loss\_function($E$, $S$, e$_{vals}$) \label{line:error}
\EndFunction
\vspace{.3cm}
\State p$_0$ $\gets$ initParams($E, F, D, n, d$) \label{line:init}
\State p $\gets$ optimizer(\Call{Eval}{}, p$_0$) \label{line:opt}
\State solF $\gets$ reconstructFunctions($n$, $d$, p) \label{line:rebuild} 
\vspace{.3cm}
\State \textbf{return } solF
\end{algorithmic}
\end{algorithm}

The algorithm returns the solutions of the differential equations in an approximate analytical form, which corresponds to the decomposition of the functions in the Chebychev basis.

\subsection{Line-by-line analysis}

In this section, we detail, line-by-line, the steps of Algorithm \ref{alg:hdes} and the related cost. For more details on some steps, we refer the reader to the appropriate section in Appendix \ref{app:runtimeDetails}.

~

\begin{itemize}
  \item In line \ref{line:obs}, the first classical pre-processing step is to build all the 
  observables that will be used later in the algorithm to evaluate the functions and their derivatives, 
  which requires the evaluation of Chebyshev polynomials at specific sample points.
   The analysis of this function is treated in Appendix 
    \ref{sub:generating_the_observables}, and the time for completion is $T_{obs} = 
    |G| \times n_s \times (2^{n-1} \times T_{cheb} + T_{diag})$, where $G$ is defined in the mentioned appendix.
  \item In line \ref{line:vqc}, we build the parametrized quantum circuits. This step
    is detailed in Appendix \ref{sub:vqc} and takes a time $T_{VQC\_init} = (2n -1) \times d \times (T_{gate\_creation} + 
T_{instruction\_adding})$.
  \item In line \ref{line:expval}, we compute the expectation value of the circuit for
    each observable. This computation is usually delegated to the quantum devices that decompose the observable and select the measurements to optimize this process. We discuss this in more detail in Appendix \ref{sub:expectation_values}, and
    this step takes a time $T_{expvalue} = n_{shots}\times (T_{VQC\_exec} + T_{measurement}) + T_{post\_estim}$.
  \item In line \ref{line:error}, we compute the loss function. This step is detailed in
    Appendix \ref{sub:error} and takes a time $T_{loss} = |E| \times n_ s \times (T_{square}+\sum_{f \in F}|H_E|T_{sum})$.
  \item In line \ref{line:init}, we initialize the angles and scaling factor for all
    functions using a uniform random number generator. The associated cost is $T_{param\_init} = n_f \times (n \times d +1)  \times T_{rand}$.
  \item In line \ref{line:opt}, we run a classical optimizer on the $\textsc{Eval}$ function. For BFGS, it has in general a complexity of $\mathcal{O}(k^2)$ \cite{NW06}, with $k$ the number of parameters to optimize, which in our case gives us $\mathcal{O}({n_f}^2 \times n^2  \times d^2)$.
  \item In line \ref{line:rebuild}, we retrieve the solution functions by running the VQC with the optimal parameters on a simulator and retrieving the probabilities of the quantum state, which give us the decomposition of the function in the Chebychev basis as an approximate analytical form. We consider this a classical post-processing step that does not affect the efficiency of the optimization algorithm.
\end{itemize}

\subsection{Scaling properties of the algorithm}

In this subsection, we detail how the resources and parameters of the algorithm
scale when we increase one specific parameter or move to more complex DEs.

~

First, when the number of variables increases, for instance when moving from ODEs
to PDEs, only the pre-processing part, the number of qubits, and the number of
samples are potentially affected. Indeed, we may need to increase the number of
sample points in the research space to also include variations in the additional
introduced variables. In the pre-processing part, it also implies the evaluation
of additional univariate Chebyshev polynomials as well as additional
multiplications (in the generation of the diagonal terms of the observable).
Furthermore, if we want to maintain a specific precision for each variable of a
function, we must multiply the number of qubits in the circuit by the number of
variables. In other words, if one needs $m$ qubits for each variable, the number
of qubits should satisfy $n = n_v \times m$, which clearly grows linearly with
the number of variables. If one decides that the number of qubits must be a fixed
value, then when the number of variables increases, the qubits (and thus the
precision) for each variable are distributed equally or not, depending on the
divisibility of $n$.

~

When the number of DEs increases, the number of solution functions naturally
increases, leading to a proportional increase in various components of the
algorithm. In fact, when we add functions, we need to add the corresponding
number of variational quantum circuits, which also introduces additional
parameters to optimize (additional angles and scaling factors). This can affect
the performance of the (BFGS) optimizer quadratically. Having more circuits to
evaluate also requires the evaluation of the expectation values (or at least the
Pauli observables that decompose the diagonal observable) for these new circuits.
However, the execution of the different circuits can be performed in parallel in
a single iteration of the hybrid loop, since the parameters of the different
circuits are independent. Finally, increasing the number of equations to solve
implies a proportional increase in summing the terms of the loss function
(and some additional boundary conditions to handle).

~

An increase in the order of the DE mainly affects the observables. First, it can
require the evaluation of additional derivatives of the Chebyshev polynomials
(if we suppose that all lower orders of derivatives appear in the DEs) in the
classical pre-processing step, but the increase is proportional to the number of
samples $n_s$. In addition, it can require the evaluation of more expectation
values, but this only affects the post-processing of such expectation values at
each iteration, because they can also be deduced from the expectation value of
the Pauli observables that are combinations of $I$ and $Z$, and thus do not
require additional circuit measurements. The same argument for expectation values
can be used when we increase the number of terms in the DEs. However, in this
latter case, the evaluation of the equations for computing the loss consequently
increases (also linearly).

~

Moving from a linear to a non-linear DE does not substantially change either the
principle or complexity of the algorithm, and this is a notable strength of our
approach. However, as mentioned in Rem.~\ref{remark:linear}, one can no longer
regroup the expectation values of the different observables into a single one,
but this only affects the post-processing of the measurements after each
iteration and depends on the number of terms (derivatives, functions, variables,
constants) in the DE.

~

When we increase the number of qubits, the number of Chebyshev polynomials in the
decomposition of the solution function increases exponentially, leading to
potentially better solution precision. Furthermore, the number of parameters to
optimize increases linearly with the number of qubits, which can affect the
performance of the (BFGS) optimizer quadratically. It also affects the
variational circuit, since for each additional qubit, we add a single-qubit
rotation gate and another CNOT gate for each layer. However, this approach is
not designed to require a large number of qubits, since with six (or seven)
qubits, we already retrieve a spectral basis of size 32 (or 64), which is
sufficient to express a large set of continuous functions.

~

When the depth of the circuit increases, more gates are executed, which linearly
increases both the execution time and the number of parameters to optimize,
quadratically affecting the performance of the optimizer. The idea behind a VQA
is to work with shallow circuits and, in our context of low-qubit circuits,
choosing an appropriate Ansatz can ensure that the required depth for good
expressivity is also low~\cite{DTYT22}.
 
~
 
An increase in the number of sample points requires the manipulation of more
observables to evaluate the function and its derivatives at those points.
Evaluating the Chebyshev polynomials at each sample point linearly increases the
pre-processing time, as well as the post-processing of the expectation values
after each iteration of the algorithm. Finally, it also increases the complexity
of computing the loss function proportionally.  

~

The final consideration is what happens when we want to increase the precision
required for the final loss function. The precision depends on several
parameters mentioned above, and to achieve better precision, one would need to
increase the number of iterations of the optimizer, increase the number of
qubits, or increase the number of sample points, depending on where the
algorithm needs help.

\section{Results}\label{sec:results}

In this section, we present the results obtained when applying our algorithm to
solve different ODEs. We first consider three examples running on emulator:
 the first is a system of two coupled linear and first-order ODEs; the second is the damped harmonic
oscillator, represented by a linear second-order ODE; and the third is a
hypoelastic deformation problem modeled by two coupled non-linear and
first-order ODEs. Finally we execute our algorithm on a real quantum backend to solve a linear first order ODE.
 
\subsection{Execution on a quantum emulator}

We run our algorithm on the ideal QLM simulator ($qlm\_35$) provided by Eviden
and give the size and depth of the circuit for each example. To generate a
final state solution, we use the BFGS~\cite{NW06}*{Sec. 6.1} optimizer to update
the circuit angles $\theta$ within our hybrid loop. The following results are
obtained by averaging over 100 random(uniform) initializations of the parameters
of the circuits. We present the results in plots showing the solution function
at each sample point as computed by both our solver and a classical
(Mathematica DSolve function) solver. For each point, we also compute and plot
the standard deviation. Finally, we present the validation scores associated
with each solution, as defined in Sec.~\ref{sec:validation}. We compute the
validation score using a set $S_v$ of 100 points, linearly spaced in the
interval $[0, 0.95]$.

\subsubsection{System of two coupled linear first-order ODEs}

\begin{figure}[!ht]
  \begin{minipage}{0.48\textwidth}
    \centering
    \includegraphics[width=1
    \linewidth]{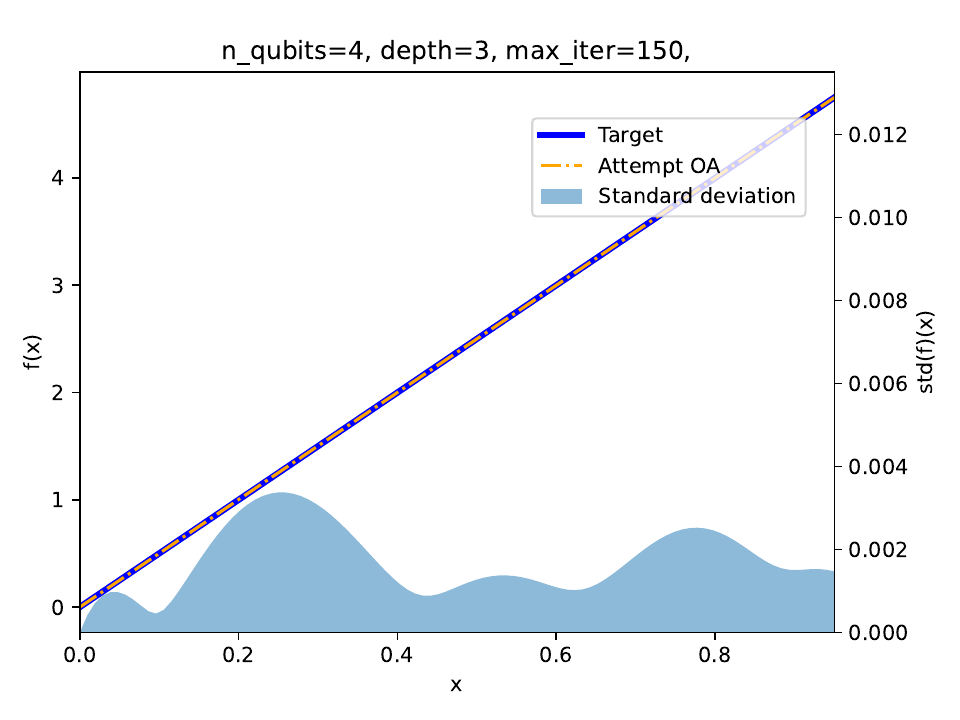}
  \end{minipage}\hfill
  \begin{minipage}{0.48\textwidth}
    \centering
    \includegraphics[width=1\linewidth]{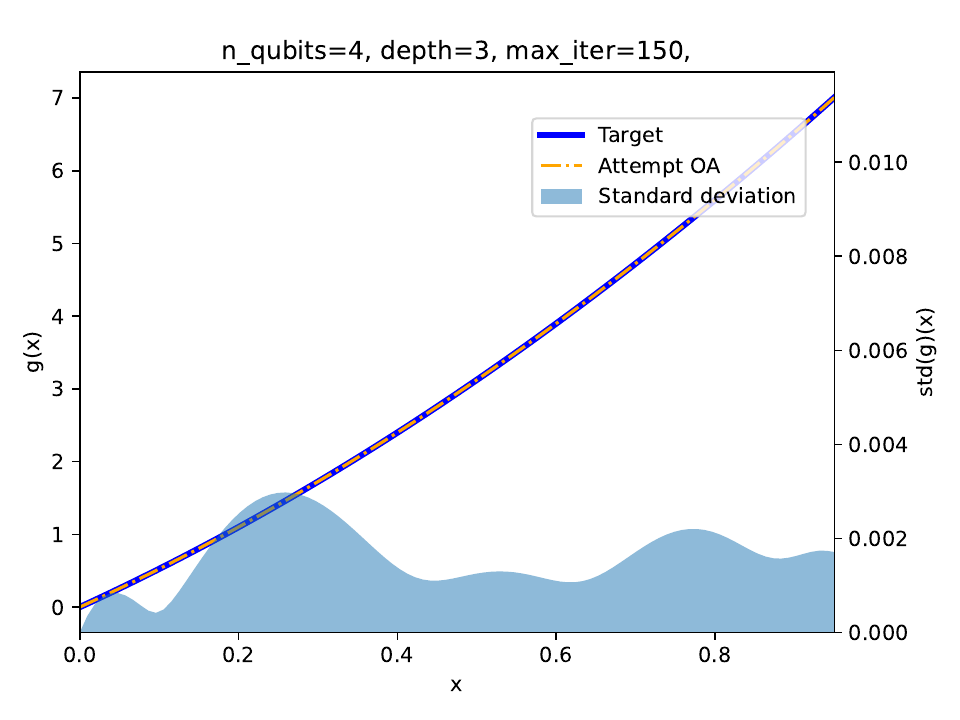}
  \end{minipage}
  \caption{The exact (blue solid) versus attempted solutions (orange dashed) for 
    $f(x)$ (left) and $g(x)$ (right) mediated over 100 attempts after 150
    iterations of the BFGS optimizer. The fit was obtained by minimizing the
    loss function over 20 points equally spaced in the $x = [0, 0.95]$ domain.
    The area at the bottom shows the standard deviation (between the attempted
    and target solution) associated with each point.}
  \label{fig:coupled_linear_1st_order}
\end{figure}

The first example we solve here is the following system of two linear coupled DEs:
\begin{equation}
  \begin{split}
    \frac{df(x)}{dx} - 5 = 0 ~,\\
    \frac{dg(x)}{dx} - f(x) - 5 = 0 ~,
  \end{split}
\end{equation}
with the boundary conditions $f(0)=0$ and $g(0)=0$. To get valuable results, we
estimated that 4 qubits, an Ansatz of depth 3, and 150 iterations with the BFGS
optimizer are sufficient. In Fig.~\ref{fig:coupled_linear_1st_order}, we compare
the attempted solution with the analytical results. We obtain a global validation
score of $V = (1.95\cdot10^{-3}, 6.20\cdot10^{-7})$ and a final loss equal to
$5.12\cdot10^{-5}$.

\subsubsection{Damped harmonic oscillator}

\begin{figure}[H]
  \centering
  \includegraphics[width=0.65\textwidth]{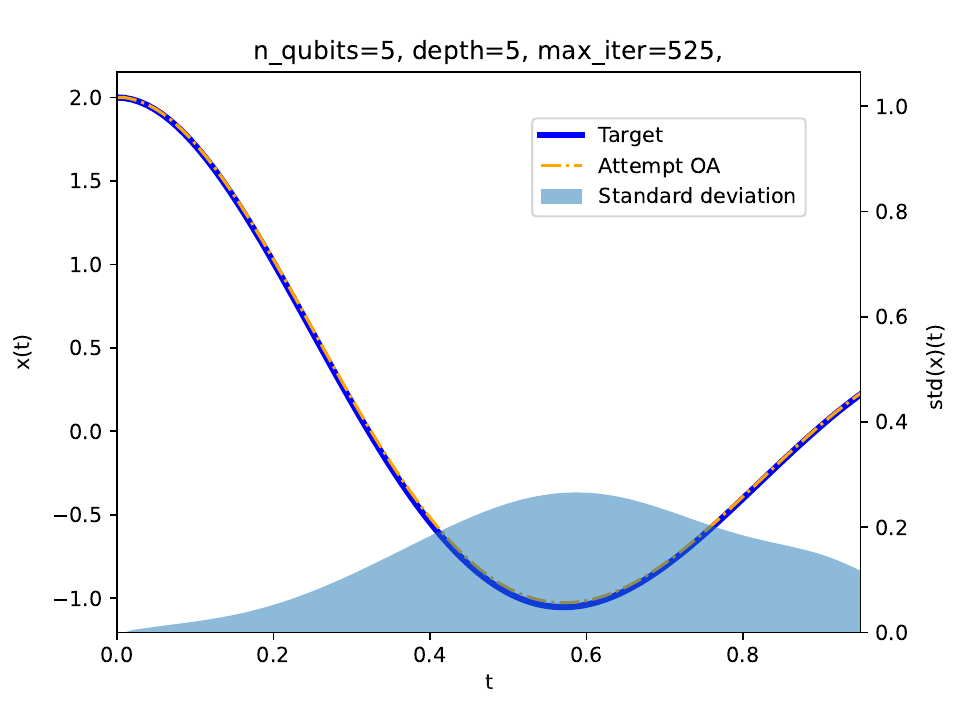}
  \caption{A comparison between the target function (blue solid) obtained 
    analytically and the attempted function of the algorithm (orange dashed). The
    area at the bottom shows the standard deviation (between the two functions)
    associated with each point.}
  \label{fig:damped_HO}
\end{figure}

The equation describing the damping of a harmonic oscillator is a linear 
second-order ODE:
\begin{equation}
  \frac{d^2 x}{dt} + 2 \zeta \omega\frac{d x}{dt} + \omega^2 x = 0 ~,
\end{equation}
with the boundary conditions $x(0)=2$ and $x'(0)=0$. Here, $\omega$ represents
the undamped angular frequency of the oscillator and $\zeta$ the damping ratio,
and these values are respectively set to $\frac{9}{8}$ and $\frac{45}{8}$. We
estimate that, to get valuable results, five qubits, an Ansatz of depth 5, and
525 iterations with a BFGS optimizer are needed.

~

In Fig.~\ref{fig:damped_HO}, we compare the attempted solution with the
analytical results. We obtain a global validation score of $V=(2.87\cdot10^{-2}, 
3.88\cdot10^{-4})$ and a final loss equal to $2.69\cdot10^{-3}$.

\subsubsection{Material deformation application}

The next example we consider is a simple material deformation problem. We consider simple and coupled DEs to describe
one-dimensional samples, such as a tensile test, where a thin strip is fixed to a
grip and pulled at the right end. One of the goals of solving DEs for static
material deformation is to compute the final displacement of the solid under a
load. This involves determining the spatial difference for each point between the
initial and deformed solids.

~

Generally, material deformation can be described in terms of forces per unit area
applied to the sample, called stress, and the resulting strain, which
characterizes the relative displacement of the body with respect to the
reference length. Mathematically, this is represented by relating a stress
tensor $\sigma_{ab}$ and a strain tensor $\epsilon_{ab}$, where $a,b=x,y,z$
denote the directions in Cartesian coordinates. This gives a set of DEs whose
form (ordinary or partial, linear or non-linear, etc.) depends on the problem
considered and approximations applied. 

~

As a proof of principle, we consider a one-dimensional case of material
deformation in the hypoelastic regime. Hypoelastic refers to a non-linear, yet
reversible, stress-strain constitutive relation. This means the deformations
(described by the strain $\epsilon_{xx}$) respond non-linearly to external
forces (the stress $\sigma_{xx}$). The governing non-linear coupled ordinary DEs
are:

\begin{equation}
  \left\{\begin{aligned}%{@{}l@ {}}
    ~\frac{du}{dx} = \epsilon_{xx}(\sigma_{xx})\\
    ~\frac{d\sigma_{xx}}{dx} +b_x = 0 
  \end{aligned}\right. ~,
\end{equation}
where $u(x)$ is the displacement along $x$ and $b_x$ is the constant change in
the stress along the $x$-direction. The stress-strain constitutive relation is given by

\begin{equation}
  \epsilon_{xx} = \frac{\sigma_{xx}}{3K} + \frac{2\epsilon_0}{\sqrt{3}}
  \left(\frac{\sigma_{xx}}{\sqrt{3} \sigma_0}\right)^n,
\end{equation}
where $b$, $K$, $\sigma_0$, and $\epsilon_0$ are material parameters defined by
the problem. We consider a thin strip of length $L$ fixed to a grip at the left
end and pulled at the right end. The corresponding boundary conditions are 
$u(0) \equiv u_0 = 0$ and $\sigma_{xx}(L)\equiv  t$.

~

\begin{figure}[!ht]
  \begin{minipage}{0.48\textwidth}
    \centering
    \includegraphics[width=1
    \linewidth]{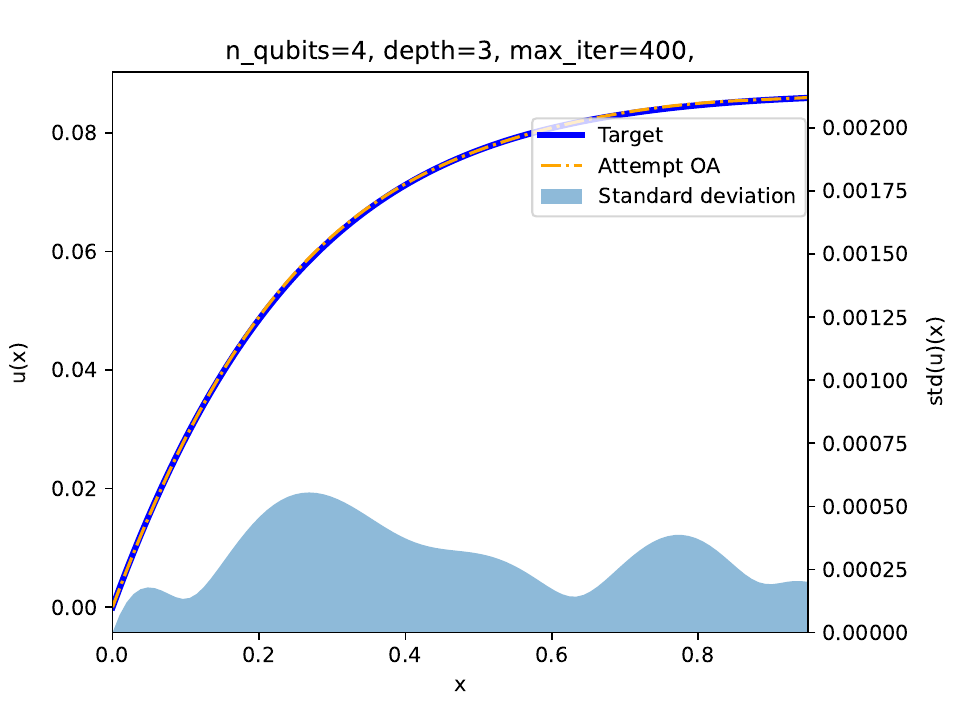}
  \end{minipage}\hfill
  \begin{minipage}{0.48\textwidth}
    \centering
    \includegraphics[width=1\linewidth]{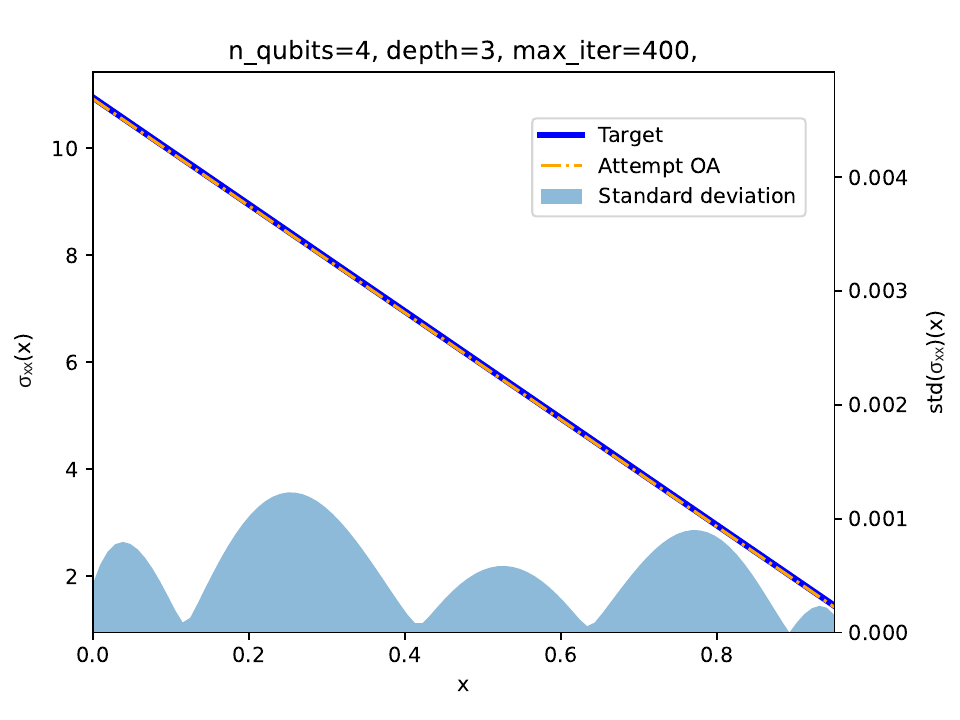}
  \end{minipage}
  \caption{The exact (blue solid) versus attempted solutions (orange dashed) for 
    the displacement (left) and stress (right) average over 100 attempts and
    after 400 iterations of the BFGS optimizer. The fit was obtained by
    minimizing the loss function over 20 points equally spaced in the $x =
    [0, 0.95]$ domain. The area at the bottom shows the standard deviation
    associated with each point.}
  \label{fig:material_deformation_1D}
\end{figure}

To simulate tensile testing (a fundamental test in material science and
engineering) of a metallic strip, we set the material parameters to $b = 10$,
$\sigma_0=5$, $\epsilon_0 = 0.1$, $n = 4$, and $K = 100$, and we use $L=0.9$ and
$t=2$ with 20 equally spaced points to sample the strip. We run our algorithm
with 4 qubits, an Ansatz of depth 3, and 400 iterations of the optimizer. In
Fig.~\ref{fig:material_deformation_1D}, we compare the attempted solution with
the analytical solutions. We obtain a global validation score of $V =
(2.59\cdot10^{-2},3.34\cdot10^{-4})$ and a final loss equal to $1.05\cdot10^{-3}$.

\subsection{Running on real quantum hardware}

The last example we consider to demonstrate the performance of our algorithm is the resolution of a simple linear first-order ordinary differential equation on a real quantum processing unit. The goal is to showcase the capabilities of the algorithm when executed on today's quantum computers on a simple enough equation. Many quantum approaches proposed in the literature for solving ODEs and PDEs fail to provide satisfying results when executed on real quantum hardware, even for relatively simple examples. In our case, we consider a canonical example described by the following ODE:

\begin{equation}
f'(x) - f(x) = 0 ~,
\end{equation}

with the boundary condition $f(0) = 1$, on the interval $[0, 0.91]$. The resolution was performed using 16 sample points uniformly distributed on the interval. We run the algorithm using 3 qubits, with a typical $R_y$ with $CNOT$ (in cascade) ansatz of depth 2, and 20 iterations of the SLSQP optimizer. At each iteration, we evaluated expectation values with respect to the generated state using 20000 shots, with an analytical computation of the gradient. The boundary condition was imposed exactly using floating boundary handling (see Appendix \ref{app:bc_shift}). The execution was made on IBM Quantum backend \emph{Brisbane} (January 2025), with Eagle r3 processor type, on which the circuit was transpiled using Qiskit 1.2.4.

~

\begin{figure}[!ht]
    \centering
    \includegraphics[width=0.65\linewidth]{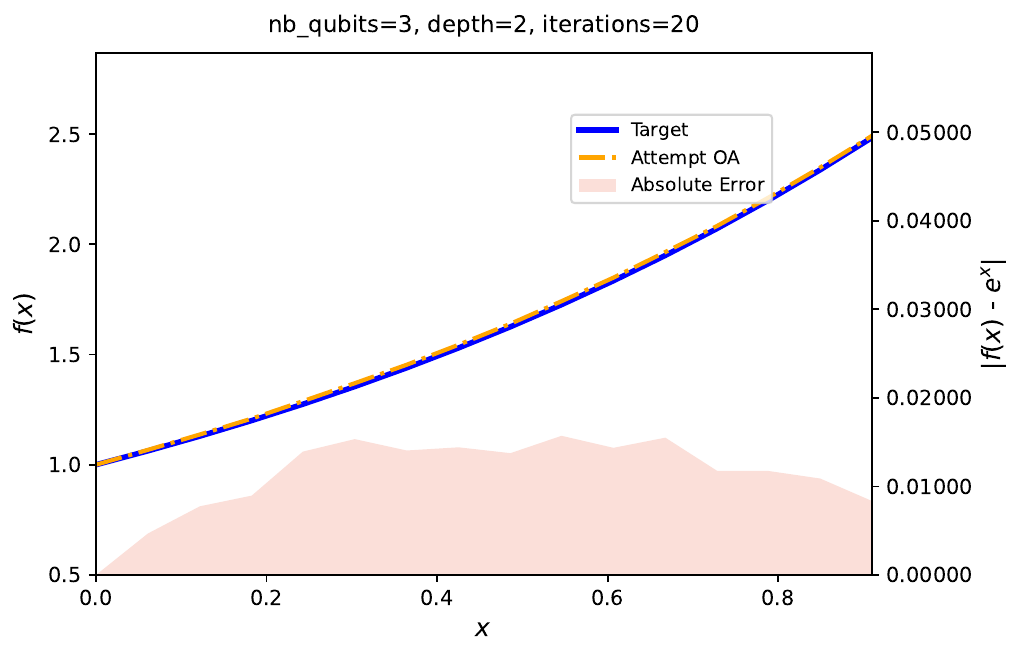}
  \caption{The exact (blue solid) versus attempted solutions (orange dashed) for 
    $f(x)$  after 20 iterations of the SLSPQ optimizer, using analytic computation of the gradient. The fit was obtained by minimizing the loss function over 16 points equally spaced in the $x = [0, 0.91]$ domain. The area at the bottom shows the absolute error (between the attempted and target solutions) associated with each sample point.}
  \label{fig:linear_1st_order}
\end{figure}

In Fig.~\ref{fig:linear_1st_order}, we compare the attempted solution with
the analytical solution. We obtain a global validation score of $V =
(1.57\cdot10^{-2},1.46\cdot10^{-4})$ and a final loss equal to $0.052$. This constitutes the first successful resolution of a differential equation on a NISQ device using the H-DES algorithm.

\section{Discussion}
\label{sec:discussions}

\subsection{Advantages of our approach}

The first advantage of our method is that the number of qubits required is
limited. First, the representativity power of our approach grows exponentially
with the number of qubits, since the number of Chebyshev polynomials doubles
when we add one qubit. With an already low number of qubits, one can represent a
large range of continuous functions. Second, the qubit requirement grows
linearly with the number of dimensions. The latter is a crucial point: since the
number of variables is usually limited by spatial and temporal dimensions
(3D$+$1D), the number of qubits needed to represent the solution function is
also limited.

~

Moreover, the number of circuits needed in this approach also remains small. In
fact, only one circuit for each iteration of the loop is required to evaluate
both a function and its derivatives. This means that the number of circuits
needed is exactly equal to the number of functions involved in a system of
partial DEs.

~

Compared to a usual polynomial interpolation in a basis of size $2^n$, we have
exponentially fewer parameters to optimize due to the exponential dimension of
the $n$-qubit Hilbert space. Indeed, with the variational quantum circuit of
$n$ qubits presented before, we only have to optimize $n\times d$ angles (with
$d$ the depth). Therefore, the number of parameters to optimize increases
linearly with the number of qubits.

~

In addition, as described before, good results using shallow circuits ($d= 3$ for
1D MD) have been obtained. The expressivity provided by the parametrized quantum
circuit allows us to find the solution without a need for deep circuits. Preliminary empirical observations on various equations seem to indicate that the depth is scaling at most linearly (sometimes logarithmically) with the number of qubits.

~

Another advantage of our method is the lack of error in the computation
of the derivatives. Since the same circuit is used to define the state
representing the function and its derivatives, no approximation is needed to
estimate them. For these reasons, we believe that the main expected advantage
provided by the algorithm is its potential to return very precise solutions with
a limited amount of quantum resources. It is also possible that this advantage
translates into a reduction in energy consumption for a fixed precision when
compared to classical solvers.

~

Finally, a general advantage is the flexibility allowed by the H-DES approach, providing a general framework for PDEs resolution, where the ansatz, the initial state, the choice of the loss function, the choice of the observable, the distribution of the sample (collocation) points (for optimization and for reconstruction of the solution) and the choice of the classical optimizer are customizable.

\subsection{Differentiation from existing algorithms}

Many quantum algorithms have been proposed in the literature for solving DEs.
However, most of the proposed methods suffer from limitations that do not allow
them to be good candidates for a general quantum DE solver. In this section, we
list the state-of-the-art quantum algorithms for solving DEs and regroup them in
Tab.~\ref{tab:validation_MD}, according to their main principles and limitations.
Most of the quantum algorithms presented are restricted in terms of types of DEs
that they can solve, and the majority of the listed methods rely on the finite
difference method which introduce approximations in the derivatives of the trial
functions. Our method has the advantage to tackle any system of partial DEs in
the NISQ era, while using an exact expression of the derivatives.

\begin{table}[!ht]
\begin{center}
\begin{tabular}{|c|c|}
\hline
Specificity & References\\
\hline
\hline
One restricted type of equation (Burger, Vlasov, ...) & \cites{GJFC,ESP19,OVKG22,LiYin2025,quanscient_press_release} \\
\hline
Solves only linear ODEs                               & \cites{Ber14, BCOW17,XWC+20,WWL+,CL20} \\
\hline
Solves only ODEs                                      & \cites{SV21,LO08,Kac06,ZMSS21} \\
\hline
Solves only linear PDEs                               & \cites{FJO21, FML17, SS19,CLO21,KDE+20,AKWL19,CJO19,LMS20,FL18,OVKG22} \\
\hline
Solves only first order  PDEs						& \cite{Pool2024} \\
\hline
Solves only polynomial DEs                            & \cite{LO08} \\
\hline
Transforms DEs into systems of linear equations       & \cites{SS19,BNWA23a,CLO21,PSG+21,KDE+20,LDG+20,CJO19,Bud21,MP16}\\
(using FEM, FDM, QFT, ...)                                 & \cites{ZMSS21,WWL+,GRG22,CL20,FL18,OVKG22,LJM+20,CLO21,LJM+20} \\
\hline
Transforms PDE into ODEs                              & \cites{Gai21,MK22b} \\
\hline
(Semi) Spectral approach                              & \cite{CL20,Schillo2025} \\
\hline
Not NISQ friendly                                     & \cites{Gai21, Ber14, CLO21,Bud21,Ste20,GRG22,SONG2025,LiYin2025,Alipanah2025} \\
\hline
Other paradigm (annealing,...)                            & \cites{SS19,PSG+21} \\
\hline
\end{tabular}
\caption{Specificities of state-of-the-art quantum algorithms for solving DEs.
  FEM stands for Finite Elements Method, while QFT stands for Quantum Fourier
  Transform.}
\label{tab:validation_MD}
\end{center}
\end{table}

We can highlight the main differences between our work and what was proposed by Kyriienko \textit{et al.} in \cite{KPE21}, who proposed a general ODE solver 
based on differentiable quantum 
circuits. In our approach, we encode the dependence on $x$ in the observables,
while Kyriienko \textit{et al.} encode it directly in the circuit using quantum
feature maps. In their approach, the evaluation of the solution function is a
non-linear process, allowing different choices of observables or quantum feature
maps, but not based on a specific encoding or decomposition of the solution
(i.e., it is not a spectral decomposition).  For evaluating the solution function or
its derivatives, we replace Chebyshev polynomials (or any other orthogonal basis) with their derivatives in the observables, 
while the approach of Kyriienko \textit{et al.} involves
reconstructing the derivatives by evaluating several circuits (changing the
quantum feature map of the circuit used to evaluate the solution function), whose
number grows exponentially with the order of the derivative. However,
Kyriienko \textit{et al.} do not discuss in detail how to solve systems of DEs,
or partial DEs and the encoding of multi-dimensional variables, for which the
cost and its scaling are also not precised.

\section{Conclusion}\label{sec:conclusion}

In this article, we have proposed an original hybrid quantum algorithm for solving
partial DEs. By encoding the solution functions in parametrized quantum circuits
and leveraging the principle of variational quantum algorithms, we have
transformed the problem of solving such equations into an optimization problem.
Using a spectral method, the algorithm is able to evaluate the solution function
and its derivatives from a single parametrized circuit by computing expectation
values of variable-dependent observables without approximating the derivatives
needed. The solver optimizes a loss function that measures how the DEs are fit on
a set of sample points. 

~

The algorithm is designed to handle any linear or non-linear DE, and is thought
to solve various use cases and multiphysics simulations. As detailed in
previous sections, its ability to be easily implemented in the case of partial
and coupled DEs demonstrates the genericity of the approach and the versatility
of our solver. Its scalability properties show that adding more complexity to the
problem by introducing non-linear terms, more variables, or more functions to the
system of DEs only linearly increases the number of resources (qubits or circuits)
and parameters to optimize. In addition, with the expressivity of the solution
function increasing exponentially with the number of qubits, and with shallow
circuits needed to encode it, the algorithm is suitable for running on current
quantum devices. We thus believe this algorithm constitutes an important step
toward toward realistic applications of hybrid quantum algorithms

~

The natural next steps will be to demonstrate the ability of our algorithm to
perform on more complicated simulations and models, moving 
towards highly dimensional, stiff, chaotic, turbulent, discontinuous and stochastic 
phenomena. Another possible improvement of the algorithm will be to
provide an initialization strategy for the circuit parameters, for instance by
exploiting the expression or properties of the DEs to start the optimization
process as close as possible to the actual solution. In the direction of having a
generic solver, one can imagine that some parameters of the solver (the number of
qubits, depth, number of iterations, number of sample points) can be
automatically chosen to satisfy the desired precision by taking into account the
type and form of the DE. Improvements to the algorithm can be made at different
stages to optimize some aspects of pre and post processing, to more efficiently
evaluate the expectation values of the observables, and to provide a
loss function less prone to barren plateaus and local minima, but also by
selecting an appropriate classical optimizer.

~

In conclusion, the hybrid quantum algorithm proposed in this paper exhibits
promising potential for practical application to various use cases despite the
current limitations inherent to quantum computing. By combining classical
computational methods with nascent quantum capabilities, this algorithm provides
an original solution for solving generic partial DEs. In the NISQ era, where
quantum computers face challenges related to stability and computational power,
the significance of this hybrid approach relies on the potential span of diverse
sectors, from material science to finance, including fluid dynamics and semiconductors,
where its deployment could help
optimize processes such as industrial design and risk assessment. This hybrid
algorithm not only serves as a bridge between theoretical quantum advancements
and industrial applicability but also underscores the current need for
accelerated integration of quantum technologies into industrial workflows. We
believe that the adoption of such algorithms will signify a critical stride
toward realizing the transformative potential of quantum computation in practical
industrial contexts.

\subsection*{Acknowledgments}

The authors want to thank Gözde Üstün, Xavier Pillet, and Ambroise Müller for
constructive discussions on a preliminary version of the algorithm. The authors
also thank Laurent Guiraud, Youcef Mohdeb, Muhammad Attallah, Roland Katz, Hugo
Bartolomei, Roman Randrianarisoa and Aoife Boyle for their valuable insights and 
feedback, which significantly enhanced the quality of this article.

\bibliographystyle{alpha}
\bibliography{library}

\newpage
\appendix

\section{Boundary conditions within the sample points}
\label{app:boundary}

In order to deal with simple boundary conditions, we first need to distinguish
between two cases: those in which the boundary condition concerns the function
$f(x)$ itself and those in which it is associated with a $q$-order derivative of
the function. In the first case, we use a method called floating boundary
conditions handling, while in the second case, exactly like for finite
difference methods, we exploit the discretization to manipulate $q+1$ points
around the coordinate $x_0$ of the boundary condition in order to make them
match the derivative at a certain point. The details of these methods will be
presented in the following.

\subsection{Floating boundary handling}\label{app:bc_shift}

This method consists of iteratively shifting the attempt solution to force
matching of the boundary conditions. Including boundary conditions through this
method allows us to solve the DE shifted to any position, consequently enlarging
the set of possible solutions. For the sake of generality, we consider the case
in which $n_{BC}$ boundary conditions at points $BC(f)=\{x_j\}_j$ need to be
applied to a function $f$, i.e.,
\begin{equation}
  \begin{split}
    f(x_1) &= k_1 ~,\\
    f(x_2) &= k_2  ~,\\
    &\vdots \\
    f(x_{n_{BC}}) &= k_{n_{BC}} ~.
  \end{split}
\end{equation} 
Let us denote as $\tilde{f}$ the attempt solution of the current iteration. We
associate each point in $BC(f)$ with a shift value as follows:
\begin{equation}
  y_j = \tilde{f}(x_j)- k_j
\end{equation}
We then perform a polynomial fit on the points $\{(x_j, y_j)\}_j$ of order 
$n_{BC}-1$, resulting in a function $\text{shift}$, such that $\text{shift}(x_j)
= y_j$ for $x_j \in BC(f)$. The new shifted attempt function is expressed as
\begin{equation}
  \tilde{f}_\text{shifted}(x) = \tilde{f}(x) - \text{shift}(x) ~.
\end{equation}

Furthermore, before evaluating the loss function at each iteration, we also have
to coherently adjust each $q$-order derivative by adding the corresponding term
$\partial^q \text{shift} / \partial x^q$. A concrete example can be found in
App.~\ref{sub:FBH-example}.

~

The main difference between this method and that including the boundary
conditions directly within the loss function is that the first takes the
boundary conditions into account exactly and with certitude, since the value of
the solution function is \textit{manually} fixed. The other solution can only
perfectly ensure the boundary condition is met if the loss function is exactly
equal to zero. Having a slight imperfection in the boundary condition fit can
lead to very different solutions, especially in a chaotic setup. However,
floating boundary conditions handling cannot be used when the boundary
conditions are more complex, and a more general interpolation 
scheme would be needed.

\subsection{Tangential approximation}

In order to impose boundary conditions on certain derivatives, we can also
consider a method called the tangential approximation or the finite difference
trick. The idea is to use the finite difference approximation of the derivatives
to impose values of the solution function on a set of close points. For example,
if a DE has a boundary condition on the first derivative of $f$ at $x_0$ of value
$k_0$, one can write the following approximation:

\begin{equation}
  \frac{\partial f}{\partial x}(x_0) = k_0 \approx \frac{f(x_1)-f(x_0)}{x_1-x_0} ~,
\end{equation}
with $x_1$ a point chosen to be \textit{close enough} to $x_0$. If we denote the 
attempt solution at the current iteration as $\tilde
{f}$, one can impose the value of
the function at $x_0$ such that:

\begin{equation}\label{eq:fd_trick}
\tilde{f}(x_0) = \tilde{f}(x_1) - k_0 (x_1-x_0) ~.
\end{equation}

By adding a point $x_1$ to the sample points $S$ of the algorithm and also
including the point $x_0$ in the evaluation of the loss function(and forcing the
value according to Eq.~\ref{eq:fd_trick}), we transform the problem of
satisfying the boundary condition in $\frac{\partial f}{\partial x}(x_0)$ into
optimizing $\tilde{f}(x_1)$ such that the finite difference is as close as
possible to the value $k_0$.

~

Once this is done, similarly to the shift presented in the previous section, we
manually replace the value of the derivative $\partial f/\partial x$ before
computing the loss function value of the attempt function $\tilde{f}(x)$. It is
worth noting that if the boundary applies to the point $x_0$, we use backward
derivative discretization; otherwise, we use forward derivative discretization.
Unlike floating boundary handling, this method is not exact, and its efficiency
increases as we increase the number of sample points or at least reduce the
distance between the points involved in the process.

\section{Examples}
\label{app:examples}

\subsection{Modeling a given function}
\label{sec:modeling}

In this section, we provide simple examples to illustrate the encoding of the
solution function in the quantum state generated by the VQC and the evaluation of
the function using observables depending on Chebyshev polynomials.

Let $f : \RR \to \RR$ be the function defined by $f(x) = 2x - 3$. Since $f$ is a
linear polynomial in $x$ of degree 1, it can be exactly expressed using the two
first Chebyshev polynomials, i.e.,

\begin{equation}
  %\begin{split}
  \text{Cheb}(0,x) = \cos(0) = 1 ~, ~~~~~\\\text{Cheb}(1,x) =  x ~.
  %\end{split}
\end{equation}

Therefore, we can express $f$ as

\begin{equation}
  f(x) = -3\cdot\text{Cheb}(0,x) + 2\cdot\text{Cheb}(1,x) ~,
\end{equation}
which we can rewrite in the form of Eq.~\ref{eq:expression_solution} as
follows:

\begin{equation}
f(x) = 5 \left( 
  \Big(0 -\frac{3}{5}\Big)\text{Cheb}(0,x) 
  + \Big(\frac{2}{5}-0\Big)\cdot\text{Cheb}(1,x) 
\right)~.
\end{equation}

We see clearly from this expression that $f$ can be modeled using the global
scaling factor $\lambda_f$ and the two-qubit quantum state $\ket{\psi_f}$ such
that:

\begin{align}
  \lambda_f    & = 5                                                                     \\
  \ket{\psi_f} & =  \frac{1}{\sqrt{5}} \left(\sqrt{2}\ket{01} + \sqrt{3}\ket{10}\right).
\end{align}

One can thus write

\begin{equation}
  f(x) = \lambda_f \langle \psi_f | O_C(x) |  \psi_f \rangle
\end{equation}

with

\begin{equation}
  O_C(x) = \begin{pmatrix}
    1 & 0 & 0  & 0  \\
    0 & x & 0  & 0  \\
    0 & 0 & -1 & 0  \\
    0 & 0 & 0  & -x
  \end{pmatrix} ~.
\end{equation}

\subsection{Floating boundary handling}
\label{sub:FBH-example}

In this section, we give a concrete example of how to handle boundary conditions
on a certain function $f: \RR \to \RR$. Let us suppose we have two boundary
conditions:
\begin{equation}
  \begin{split}
    f(x_0) &= a ~,  \\
    f(x_1) &= b ~.
  \end{split}
\end{equation} 

At each step of the optimization process, we shift the attempt solution 
$\tilde{f}(x)$ by constructing a straight line shift function using the following
values:
\begin{equation}
  \begin{split}
    \text{shift}(x_0) &= \tilde{f}(x_0)-f(x_0) = \tilde{f}(x_0) - a ~, \\
    \text{shift}(x_1) &= \tilde{f}(x_1)-f(x_1) = \tilde{f}(x_0) - b ~.
  \end{split}
\end{equation}
Then, the new attempt function is:
\begin{equation}
  \tilde{f}_{\text{shift}}(x)= \tilde{f}(x) - \text{shift}(x).
\end{equation}
In this specific example, since we are dealing with a linear shift, we only have
to adjust the first-order derivative $\partial f / \partial x$ as follows:
\begin{equation}
\frac{\partial \tilde{f}_{\text{shift}}(x)}{\partial x} = 
  \frac{\partial \tilde{f}(x)}{\partial x} - m,
\end{equation}
where $m$ is the angular coefficient associated with $\text{shift}(x)$.

\subsection{Simple example in the multivariate case}
\label{sec:multivariate_example}

Let $\ket{\psi_g} \in (\mathbb{C}^2)^{\otimes 6}$ be a six-qubit state and let
$\lambda_g= 7$ be a scalar used to model a bivariate solution function $g :
  \RR\times\RR \to \RR$, such that :

$$\ket{\psi_g} =  \frac{1}{\sqrt{14}} \left( 2\ket{001000} + 2\ket{100010} +
  \sqrt{6}\ket{010111} \right) ~,$$
with

$$g(x,y) = \lambda_g \langle \psi_g | O_C(x,y) |  \psi_g \rangle$$

We choose to encode the $x$-dependence with three qubits and the $y$-dependence
with two qubits. Therefore, we can divide any basis state $\ket{i}$ of six-qubit
systems as follows:

$$\ket{i} = \underbrace{\ket{i_0}}_{\text{sign}} \underbrace{\ket{i_1 i_2
      i_3}}_{\ket{L_1}} \underbrace{\ket{i_4 i_5}}_{\ket{L_2}} ~,$$
with $\textrm{bin}(L_1) = i_1 i_2 i_3$ and $\textrm{bin}(L_2) = i_4 i_5$. If we
take for instance the first basis state appearing in $\ket{\psi_g}$, we can
associate it with the following multivariate Chebyshev polynomial after
computation of the expectation value of the diagonal observable $O_C(x,y)$:

$$\underbrace{\ket{0}}_{+} \underbrace{\ket{010}}_{2} \underbrace{\ket{00}}_{0}
  ~ \xrightarrow{\langle O_C(x,y) \rangle} ~ \text{Cheb}(2,x)\text{Cheb}(0,y) =
  2x^2-1 ~,$$
with

$$O_C(x,y) = \text{diag}\Big( \text{Cheb}(0,x,y), \text{Cheb}(1,x,y), \dots,
  -\text{Cheb}(30,x,y), -\text{Cheb}(31,x,y) \Big) ~,$$
which is equal to:
\begin{equation*}
  \begin{split}
    O_C(x,y) = \text{diag}\Big(
    \text{Cheb}(0,x)\text{Cheb}(0,y), \text{Cheb}(0,x)\text{Cheb}(1,y), \dots, \\
    -\text{Cheb}(7,x)\text{Cheb}(2,y), -\text{Cheb}(7,x)\text{Cheb}(3,y)
    \Big) ~.
  \end{split}
\end{equation*}

Therefore, we can express the solution function $g$ as follows:

$$g(x,y) = \frac{7}{14} \Big( 4\cdot\text{Cheb}(2,x)\text{Cheb}(0,y)
  -4\cdot\text{Cheb}(0,x)\text{Cheb}(2,y) +6\cdot\text{Cheb}(5,x)\text{Cheb}(3,y)
  \Big) ~.$$

We can rewrite the univariate polynomials to obtain:

$$g(x,y)=\frac{1}{2}\Big(4(2x^2-1) - 4(2y^2-1) +
  6(16x^5-20x^3+5x)(4y^3-3y)\Big)~.$$

After some simplifications, we conclude that:

\begin{equation}
  g(x,y)=4x^2 - 4y^2 - 45xy + 180x^3y + 60xy^3 - 144x^5y - 240x^3y^3 + 192x^5y^3~.
\end{equation}

\section{Run time analysis details} \label{app:runtimeDetails}
\label{sec:algorithm_sections_details}

This section details the time taken to run various parts of the algorithm.

\subsection{Generating the observables}
\label{sub:generating_the_observables}

In this subsection, we study the function called in line \ref{line:obs} of 
Alg.~\ref{alg:hdes}.

\begin{algorithm}[H]
\caption{Observables generation function} \label{alg:obs}
\begin{algorithmic}[1]
\Require $G_F$ : the set of orders of derivatives involved in $E$ for each function in $F$
\Require $D$ : the set of domains of definition for each function in $F$
\Require $n$ : the number of qubits
\Require $n_s$ : the number of samples
\Ensure the list of observables capable of computing the value of any 
  function's derivative at any sample point, where the function is encoded in
  the output state of our VQA \vspace{.5cm}

\Function{GenerateObservables}{$G_F, D, n, n_s$}
  \State $G \gets \cup_{f\in F} G_F(f)$ \label{line:union}
  \State $S \gets$ generateSamples($n_s$, $D$) \label{line:samples}
  \State listObs $\gets$ [~]
  \For {$g$ \textbf{in} $G$} \label{line:iter:deriv}
    \State $\text{order}_g$ $\gets$ [~]
    \For {$x_s$ \textbf{in} $S$} \label{label:iter:points}
      \State $\text{C}$ $\gets$ [~]
      \For {i \textbf{from} 0 \textbf{to} $2^{n-1}-1$} \label{line:iter:cheb}
        \State $\text{C}$.append(Cheb($i$, $x_s$, $g$)) \label{line:cheb}
      \EndFor
      \State observable $\gets$ diag(C, $-$C) \label{line:diag}
      \State $\text{order}_g$.append(observable)
    \EndFor
    \State listObs.append($\text{order}_g$)
  \EndFor

  \State \textbf{return } listObs, $S$
\EndFunction
\end{algorithmic}
\end{algorithm}

$$T_{obs} = T_{union} + T_{samples} + T_{der\_iters} ~.$$

In Alg.~\ref{alg:obs}:
\begin{itemize}
  \item In line \ref{line:union}, we deduce all the derivation levels in the DEs in time
    $T_{union}$. This step is just a union of sets, so we assume that $T_{union}
    \ll T_{obs}$.
  \item In line \ref{line:samples}, we compute all the sample points used for the
    evaluation in $T_{samples}$. This step consists of the creation of a list of
    regularly spaced points, and we assume that $T_{union} \ll T_{samples}$.
  \item In line \ref{line:iter:deriv}, we iterate over each order of derivation (0, 1,
    ...) in $T_{der\_iters} = |G|\times T_{samp\_iter}$.
  \item In line \ref{label:iter:points}, we iterate over each sample point in 
    $T_{samp\_iters} = n_s\times (T_{order\_iter} + T_{diag})$.
  \item In line \ref{line:iter:cheb}, we iterate over each Chebyshev polynomial's order
    and compute the corresponding value for the Chebyshev monomial in 
    $T_{order\_iter} = 2^{n-1} \times T_{cheb}$. We can remark here that it is 
    possible to enhance this part by not computing the evaluation of 
    Chebyshev polynomials that are used for several orders twice.
  \item In line \ref{line:cheb}, we compute the value of the Chebyshev monomial 
    for a given order, sample value, and derivative order in $T_{cheb}$.
  \item In line \ref{line:diag}, we instantiate the diagonal observable, with the 
    elements on the diagonal being the input of the function. This can be done by
    instantiating a matrix or by using the Pauli string decomposition of the
    observable given in Eq.~\ref{eq:observable_decom}. This step takes a
    time $T_{diag}$ (which is linearly dependent on the size of the input).
\end{itemize}

The overall time is then:
\begin{equation}
T_{obs} = |G| \times n_s \times (2^{n-1} \times T_{cheb} + T_{diag})
\end{equation}

\textbf{Some theory:} The observable in this algorithm can be written as:\\

\begin{gather}\label{eq:observable_decom}
O_C(x) = \sigma_z \otimes \left(\sum_{i=0}^{2^{n-1}-1} \text{Cheb}(i,x) 
  \ketbra{i}{i} \right)\nonumber\\
       = \begin{pmatrix}
\text{Cheb}(0,x)  &        &                          &                   &        &                           \\
                  & \ddots &                          &                   &        & \text{\huge0}             \\
                  &        & \text{Cheb}(2^{n-1}-1,x) &                   &        &                           \\
                  &        &                          & -\text{Cheb}(0,x) &        &                           \\
\text{\huge0}     &        &                          &                   & \ddots &                           \\
                  &        &                          &                   &        & -\text{Cheb}(2^{n-1}-1,x) \\
\end{pmatrix} ~.
\end{gather}

In order to generate the full matrix, we construct the diagonal of the matrix
using the tensor product of linear combinations of the Pauli matrix $Z$ and the
identity $I = I_2$, as well as ${\rm bin}(i)$, the binary representation of $i$:

\begin{equation}
\ketbra{i}{i} = \bigotimes_{k\in {\rm bin}(i)}\frac{1}{2} (I + (\shortminus 1)^k Z) ~.
\end{equation}

Then, using the tensor products, it takes $n(T_{tensor,2\times2} + T_
{sum,2\times2})$ to construct $O_C(x)$ (per given point $x$), with $T_
{tensor,2\times2}$ the time required to compute the tensor product of a matrix with a
$2\times2$ matrix, and $T_{sum,2\times2}$ the time to add or subtract two
$2\times2$ matrices.

\subsection{Variational quantum circuit (VQC)}
\label{sub:vqc}

We use the so-called hardware-efficient Ansatz \cite{CAB+21} to construct the
state encoding the solution function. Each layer contains parameterized
single-qubit rotations for all the qubits, which are then entangled by CNOT
gates in a \textit{braided} (or alternating) fashion.

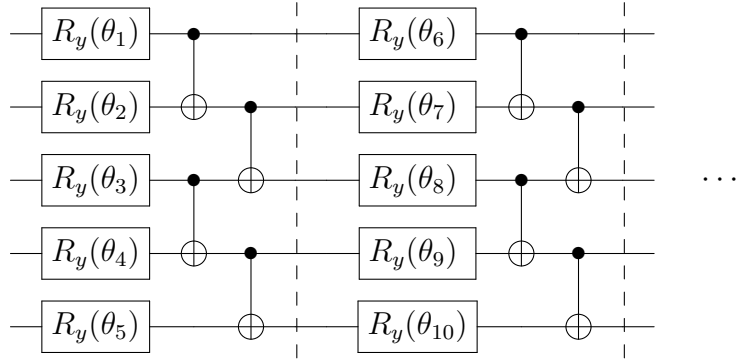
\begin{figure*}[!ht]
$$
\Qcircuit @C=1em @R=.7em {
&\gate{R_y(\theta_1)} &\ctrl{1}  &\qw      & \qw \barrier{4} & \qw & \gate{R_y(\theta_6)\;}  &\ctrl{1}  &\qw      & \qw \barrier{4} & \qw &  \\
&\gate{R_y(\theta_2)} &\targ     &\ctrl{1} & \qw             & \qw & \gate{R_y(\theta_7)\;}  &\targ     &\ctrl{1} & \qw             & \qw &  \\
&\gate{R_y(\theta_3)} &\ctrl{1}  &\targ    & \qw             & \qw & \gate{R_y(\theta_8)\;}  &\ctrl{1}  &\targ    & \qw             & \qw & \rstick{\cdots} \\
&\gate{R_y(\theta_4)} &\targ     &\ctrl{1} & \qw             & \qw & \gate{R_y(\theta_9)\;}  &\targ     &\ctrl{1} & \qw             & \qw &  \\
&\gate{R_y(\theta_5)} &\qw       &\targ    & \qw             & \qw & \gate{R_y(\theta_{10})} &\qw       &\targ    & \qw             & \qw &  
}
$$
\caption{An example of our VQC with $n=5$ qubits and depth $d=2$.}
\label{fig:vqc}
\end{figure*}

There are two types of times attached to the VQC: the initialization of the circuit
and the execution of the circuit. The initialization takes a time 
\begin{equation} \label{eq:vqc-init}
T_{VQC\_init} = (2n -1) \times d \times (T_{gate\_creation} + 
T_{instruction\_adding})
\end{equation}
and the execution of the circuit takes a time
\begin{equation} \label{eq:vqc-exec}
T_{VQC\_exec} = 3 \times d \times T_{gate\_exec}.
\end{equation}

One can remark that, even in the presence of several functions in the
differential equations, all the circuits can be run in parallel. When the number
of qubits and the depth are the same for all functions, we can also imagine only
using a single circuit that is initialized once and just replacing
the angles to evaluate one function or another.

\subsection{Expectation values}
\label{sub:expectation_values}

In order to determine the value of the solution $f$ at a given point $x$, we
compute the expectation value with respect to the diagonal observable $O_C
(x)$ discussed above as follows:

\begin{equation}
  f_{\theta}(x) = \lambda ~ \expval{O_C(x)}{\psi(\theta)}.
\end{equation}

Though one has to deal only with diagonal matrices, evaluating the expectation
values of several observables constitutes the costly step of the quantum part of
the algorithm. The method of expressing the elements of the matrices is convenient 
since evaluating the expectation values of Pauli strings through a quantum
computer is practical, especially if the Pauli strings commute, since in that
case they can be measured simultaneously~\cite{KRP22}. The upper bound of the
number of such strings \textit{per observable} is $2^{n-1}$, with $n$ the number of
qubits. So in the worst case scenario, in which no strings commute, one has
$T_{eval} = 2^{n-1} T_{measurement}$. 

~

But in our case, our Pauli strings have a very specific form: they all start with
a Pauli $Z$ gate and can then take all combinations of Pauli $I$ and $Z$. In
that case, all the Pauli operators used to decompose our diagonal observables
commute. Therefore, one can imagine many protocols and methods
for efficiently evaluating the expectation values of such observables, even in
the presence of noise \cites{ZNCA,KSK+22,HKP21,CG23a,HBRM20,KIK+22,RB20,HKP20,KG22}.
Once we know all the expectation values of the Pauli observables, we can
reconstruct all expectation values of all diagonal observables (for all samples,
for evaluating solution functions or their derivatives) in parallel with classical
post-processing.

~

The evaluation of the expectation values consists of running the
circuit and carrying out the right measurements for each shot, then
post-processing them to reconstruct all desired expectation values, giving

\begin{equation} \label{eq:single-ev-time}
T_{expvalue} = n_{shots}\times (T_{VQC\_exec} + T_{measurement}) + T_{post\_estim}.
\end{equation}

\subsection{Loss function (error)}
\label{sub:error}

The value we want to minimize is:
\begin{equation}
L^{\rm{diff}}(\theta) = 
  \frac{1}{n_s} \sum_{e \in E} \sum_{x_s \in S} e(x_s)^2 ~,
\end{equation}
where the notation is the same as that defined in Sec.~\ref{sub:loss}.

~

We omit the term in the loss that concerns the boundary conditions and apply
floating boundary conditions handling. Recall that we have:
\begin{itemize}
  \item $\theta$: the set of angles parametrizing our solution,
  \item $E$: the set of differential equations,
  \item $n_s$: the number of samples,
  \item $F$: the set of functions,
  \item $n_f$: the number of functions involved in the equations, 
\end{itemize}

To compute this sum, we iterate over:
\begin{enumerate}
  \item the set of differential equations $E$,
  \item the set of samples ($n_s$ elements),
  \item the set $H_E$ of all solution functions and their derivatives 
    appearing in the differential equations.
\end{enumerate}

Given this, the time to compute this loss value in the worst case (for all
functions, with the function and its derivatives appearing in each equation) would
be: 
\begin{equation} \label{eq:error}
T_{loss} = |E| \times n_ s \times (T_{square}+\sum_{f \in F}|H_E|T_{sum}),
\end{equation}
where $T_{square}$ is the time required to square a floating point number and $T_{sum}$ is
the time needed to sum two floating point numbers.

\section{Performance profile}\label{app:performance_profile}

Performance profiles, as described in~\cite{DM02}, are a fairly common tool
in optimization when it comes to comparing the performance of different
algorithms, software, or methods for solving a problem. They allow for assessing
both the speed and accuracy of an algorithm in relation to the proposed
solutions. They can also be used to compare the performance of a given solver
with different parameters and settings.

~

We denote the set of solvers to compare as $\mathcal{S}$ and the set of problems
on which the solvers will be compared as $\mathcal {P}$. In the context of this
paper, the problems to solve correspond to different sets of partial DEs. One
then usually needs to choose a criterion to measure the performance or quality of
the solver. It can be the number of evaluations of the solution function, the
final value of the loss function, the precision of the result, or the number of
iterations needed for a given precision. In our case, the validation score we
defined previously is used as the measure of performance for a given solver.

~

We thus define the quantity $V_{p,sol}$, which is the validation score of the
solver $sol \in \mathcal{S}$ for solving the problem $p \in \mathcal{P}$. In
order to be able to compare several solvers, we need to have a common basis for
comparison. As detailed in~\cite{DM02}, for a given problem $p$, we compare the
performance of each solver with the best performance over $\mathcal{S}$. We then
define the performance ratio $r_{p,sol}$ as follows:

\begin{equation}
  r_{p,sol} = \frac{V_{p,sol}}{\min\limits_{sol \in \mathcal{S}}  V_{p,sol} } ~.
\end{equation}

Therefore, for a given problem $p$, we define the performance ratio as the
validation score of the solver for this particular problem divided by the best
validation score (the minimum) among all other solvers, still for the same
problem. We determine that the solver could not solve the problem
if the validation score is above the acceptance threshold. If a solver is not
able to solve a problem, the performance ratio is set to a maximum value $r_M$.

~

We can now define the global performance of a solver $sol$ for all problems in
$\mathcal{P}$ as the cumulative distribution function

\begin{equation}
\rho_{sol}(\tau) = 
  \frac
    {\text{card}\Big(\{p \in \mathcal{P} ~|~ r_{p,sol}\leq \tau\}\Big)}
    {\text{card}(\mathcal{P})} ~,
\end{equation}
where $\rho_{sol}(\tau)$ is the probability for the solver $sol$ that a
performance ratio $r_{p,sol}$ is within a factor $\tau \in \RR$ of the best
possible ratio (which is 1). The function $\rho_{sol}$ can be seen as the
(cumulative) distribution function for the performance ratio. To obtain the
performance graph of the solver $sol$, we will represent the strictly increasing
function $\rho_{sol}: \RR \to [0,1]$ as a function of $\tau$.

~

The value $\rho_{sol}(1)$ is the probability that the solver $sol$ has a better
validation score than all the other solvers. Therefore, if we are interested in
precision, we can only compare the values of $\rho_{sol}(1)$ for all solvers in
$\mathcal{S}$.

~

Furthermore, we recall that any performance ratio takes a value within $[1,r_M]$, 
and that $r_{p,sol} = r_M$ only if the problem $p$ was not solved by $sol$. We
can deduce that $\rho_{sol}(r_M) = 1$ and that the quantity

\begin{equation}
  \rho_{sol}^* = \lim_{\tau \rightarrow {r_M}^{-}} \rho_{sol}(\tau)
\end{equation}
is the probability that $sol$ is able to solve a problem. Thus, if we are only
interested in solvers that have the highest probability of success, we
compare values of $\rho_{sol}^*$ for all solvers in $\mathcal{S}$ and keep
that with the highest value. The value of $\rho_{sol}^*$ can be easily
observed in a performance profile graph because the value of $\rho_ {sol}$
stagnates for high values of $\tau$.

\begin{figure}[H]
  \centering
  \includegraphics[width=0.65\textwidth]{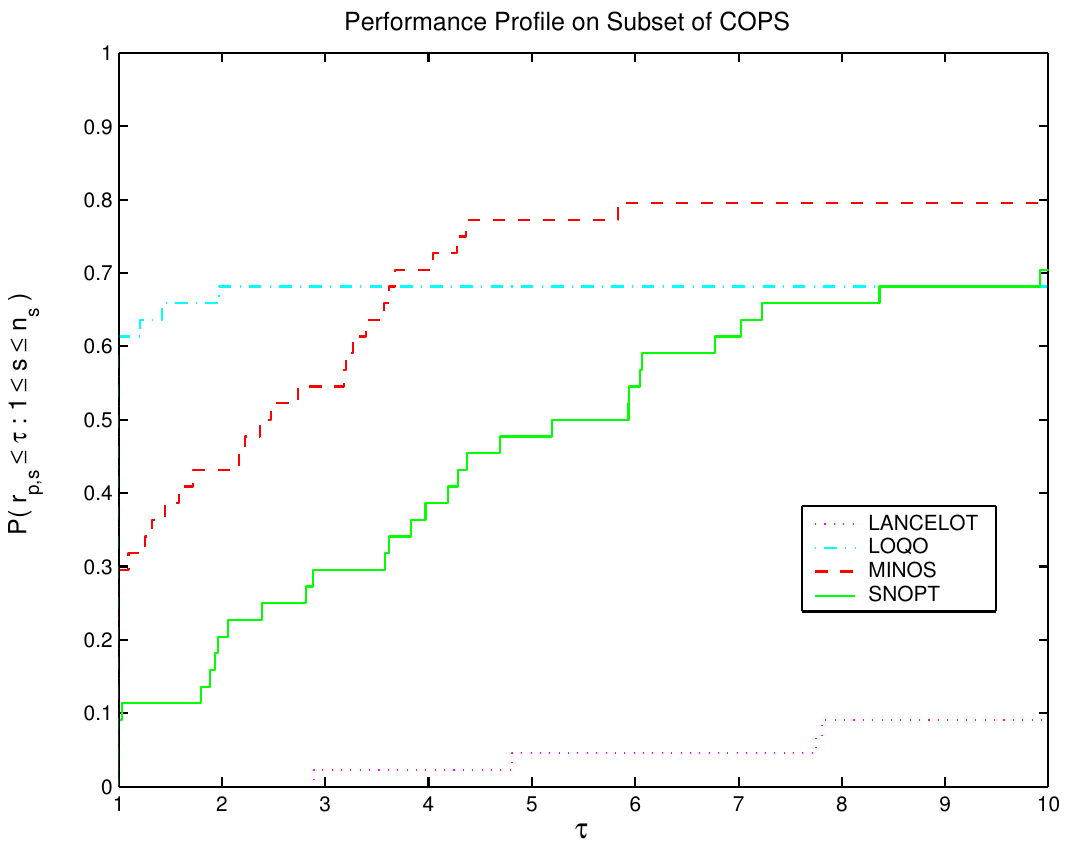}
  \caption{Example of a performance profile graph from~\cite{DM02}. The highest
    value of $\rho_{sol}(1)$ is here achieved by LOQO (0.61), while the highest
    value of $\rho_{sol}^*$ is apparently achieved by MINOS (0.8) for $\tau \in
      [1,10]$. }
  \label{fig:perf_profile}
\end{figure}

\end{document}